\documentclass[aps,pre,superscriptaddress,reprint,
amsmath,amssymb,floatfix]{revtex4-1}
\usepackage{graphicx}
\usepackage{bm}
\usepackage{siunitx}

\newcommand{\AVE}[1]{\ensuremath{\langle {#1} \rangle}}
\newcommand{\ABS}[1]{\ensuremath{\lvert {#1} \rvert}}
\newcommand{\dpone}[2]{\ensuremath{\displaystyle\frac{\partial {#1}}{\partial {#2}}}}
\newcommand{\dptwo}[2]{\ensuremath{\displaystyle\frac{\partial^2 {#1}}{\partial {#2}^2}}}

\newcommand{\bC}{\ensuremath{\bm{C}}}

\newcommand{\bF}{\ensuremath{\bm{F}}}

\newcommand{\bI}{\ensuremath{\bm{I}}}

\newcommand{\bL}{\ensuremath{\bm{L}}}
\newcommand{\bM}{\ensuremath{\bm{M}}}

\newcommand{\bQ}{\ensuremath{\bm{Q}}}

\newcommand{\bU}{\ensuremath{\bm{U}}}

\newcommand{\bX}{\ensuremath{\bm{X}}}

\newcommand{\bg}{\ensuremath{\bm{g}}}

\newcommand{\bj}{\ensuremath{\bm{j}}}

\newcommand{\bmm}{\ensuremath{\bm{m}}}
\newcommand{\bn}{\ensuremath{\bm{n}}}

\newcommand{\bq}{\ensuremath{\bm{q}}}
\newcommand{\br}{\ensuremath{\bm{r}}}

\newcommand{\bx}{\ensuremath{\bm{x}}}

\newcommand{\bsigma}{\ensuremath{\bm{\sigma}}}

\newcommand{\nablas}{\ensuremath{\hat{\nabla}_*}}

\newcommand{\hnabla}{\hat{\nabla}}
\newcommand{\hr}{\hat{r}}

\newcommand{\hkappa}{\hat{\kappa}}
\newcommand{\ntop}{\ensuremath{n_{top}}}

\begin{document}
	\title{The curved kinetic boundary layer of active matter}
	
	\author{Wen Yan}
	\email[]{wyan@flatironinstitute.org}
	\affiliation{Department of Mechanical \& Civil Engineering, 
		Division of Engineering \& Applied Science, 
		California Institute of Technology}
	\altaffiliation{Current Address: Center for Computational Biology, Flatiron Institute, Simons
		Foundation}
	\author{John F. Brady}
	\email[]{jfbrady@caltech.edu}
	\affiliation{Division of Chemistry \& Chemical Engineering 
		and Division of Engineering \& Applied Science 
		California Institute of Technology}
	
	\date{\today}
	
	\begin{abstract}
		The finite reorient-time of swimmers leads to a finite run length $\ell$ and the kinetic accumulation boundary layer on the microscopic length scale $\delta$ on a non-penetrating wall. That boundary layer is the microscopic origin of the swim pressure, and is impacted by the geometry of the boundary [Yan \& Brady, \textit{J. Fluid. Mech.}, 2015, \textbf{785}, R1]. In this work we extend the analysis to analytically solve the boundary layer on an arbitrary-shaped body distorted by the local mean curvature. The solution gives the swim pressure distribution and the total force (torque) on an arbitrarily shaped body immersed in swimmers, with a general scaling of the curvature effect $\Pi^{swim}\sim\lambda\delta^2/L$.
	\end{abstract}

	\maketitle
	
	
	
	


\section{introduction}
The accumulation of swimming micro-organisms adjacent to a boundary is often explained as a result of hydrodynamic interactions with the boundary.\cite{Berke2008} The accumulation may also result from steric effects where an elongated bacterium cannot freely swim on a surface due to geometric constraints with the wall.\cite{Li2009,Li2011} The accumulation and the resulting wall-swimmer interactions show very interesting behavior, including rheotaxis\cite{Uspal2015,Kaya2012} and circular motion.\cite{Lauga2006} Also, because of this interaction a macroscopic body may be able to harvest energy from a bacterial solution and achieve net motion simply due to its asymmetric shape.\cite{Kaiser2014}  

In fact,  active swimmers need not be elongated nor interact hydrodynamically with a surface to exhibit this boundary accumulation.\cite{SwimForce2015,Ezhilan2015} It may be simply due to the fact that when a swimmer hits a wall, it  maintains its swim orientation towards the wall for a finite time $\tau_R$.  In contrast, swimmers pointing away from the wall simply swim away.  Thus, there is an accumulation of swimmers adjacent to a surface with a net orientation towards the boundary.  This phenomena can be understood and quantitatively described by the  Active Brownian Particle (ABP) model. Each ABP propels itself at a fixed swim velocity $\bU^{swim} = U_0\bq$, where $U_0$ is its (constant) swim speed, and its orientation, $\bq$, is subject to rotational Brownian diffusion $D_R=1/\tau_R$, which sets the reorientation time $\tau_R$.  In general, ABPs are also  subject to translational Brownian motion characterized by a  diffusivity $D_T = k_BT/\zeta$, where $\zeta$ is the (isotropic) drag coefficient and $k_BT$ is the thermal energy.

When an ABP is stuck on a wall, it transmits to the wall its swim  force $-\zeta U_0\bq\cdot\bn$ because it cannot cross the boundary; here $\bn$ is the wall surface normal vector. That force accumulates over time and space, and the net effect constitutes a pressure on the wall. 
That simple process is precisely the microscopic origin of the `swim pressure'.\cite{Pressure2014} In the absence of hydrodynamic interactions for a `hard' wall, \citet{ForceBoundary2015} showed that the pressure of ABPs arises from a natural extension of the pressure of Passive Brownian Particles (PBP): $\Pi_{wall} = k_BT n_{wall}$, where $n_{wall}$ is the number density of particles at the wall.  Due to this kinetic boundary accumulation  $n_{wall} = n^\infty(1 + \frac{1}{6}(\ell/\delta)^2)$, where $n^\infty$ is the constant number density far from the wall, $\ell = U_0 \tau_R$ is the run length of the active particles and $\delta=\sqrt{D_T\tau_R}$ is a microscopic length.  The ratio $\frac{1}{6}(\ell/\delta)^2 = k_sT_s/k_BT = D^{swim}/D_T$, with the `swim' diffusivity $D^{swim} = U_0^2 \tau_R /6$ (in 3D).   Thus, the pressure at the wall $\Pi_{wall} = k_BT n_{wall} = (k_BT + k_sT_s) n^\infty$ is the sum of the passive Brownian osmotic pressure, $\Pi^{osm} = k_BT n^\infty = \zeta D_T n^\infty$, and the swim pressure, $\Pi^{swim} = k_sT_s n^\infty =  \zeta D^{swim} n^\infty = \zeta (U_0^2 \tau_R /6) n^\infty$.   Here, $k_sT_s = \zeta U_0^2\tau_R/6$ defines the `activity' of the swimmers and in certain situations plays a role analogous to the thermal energy $k_BT$ of passive particles.\cite{takatori_forces_2016}

\citet{ForceBoundary2015} showed for ABPs that this wall accumulation occurs over a microscopic length $\lambda^{-1} = \delta/\sqrt{2[1 + \frac{1}{6}(\ell/\delta)^2]}$, leading to a concentration boundary layer adjacent to the surface.  The boundary concentration, $n_{wall}$, can be determined for various geometries by a moment expansion method,\cite{Saintillan2015} and including high order moments is usually not necessary.\cite{ForceBoundary2015} The results naturally encompass the singular limit $\delta\to 0$, which applies when there is no translational Brownian motion, $D_T\to 0$.\cite{Ezhilan2015,FilyConfinedSwimmers2014}

It has been shown that the boundary layer is significantly impacted by the boundary curvature.\cite{ForceBoundary2015,FilyConfinedSwimmers2014,Smallenburg2015} Therefore, by careful design of the variation of the boundary curvature, an asymmetric macroscopic body immersed in ABPs can achieve a net force\cite{ForceBoundary2015} and therefore motion. However, a general theoretical understanding of the role of boundary curvature is lacking. And, importantly, the connection between the inner boundary-layer structure and the `outer' bulk concentration field is unknown. 

In this work, we extend the analysis of \citet{ForceBoundary2015}  to an arbitrary shaped smooth body with continuous principal curvatures when the body size $L$ is much larger than both the microscopic length $\delta$ and the run length $\ell$, i.e. $\delta, \ell \ll L$. In Section 2 we start from the Smoluchowski equation for ABPs with its moment-hierarchy expansion, and discuss the separation of scales in the governing equation by presenting a general solution to the Smoluchowski equation. 

In Section 3 we perform a boundary-layer analysis of the Smoluchowski equation to reveal the leading order effects of  curvature. In the boundary layer, the inner solution is found by building a local curvilinear coordinate system with the principal directions of curvature. We then match the inner solution to the outer solution to calculate the net force on an arbitrary shaped body. At the leading, zeroth order, curvature does not enter and the swim pressure everywhere on the surface of the body is simply $n^\infty k_sT_s$. In this is the `continuum limit' there is no net force, just like an arbitrary shaped body in the atmosphere feels a constant pressure everywhere and therefore experiences no net force.

In the leading order effects of curvature, only the mean curvature appears. Therefore without considering variations of curvature along the body surface, there is no tangential flux of swimmers in the boundary layer, and by continuity of the swimmer number density, the outer solution has a no flux boundary condition and is found to be a constant $n_{out}=n^\infty$. At  this order the effects of curvature on the swim pressure is found analytically to scale as $(J_S\lambda\delta^2/L) n^\infty k_sT_s$ everywhere on the body when $\lambda\delta^2/L\ll1$, where $J_S$ is twice the (non-dimensionalized) mean curvature. The leading order net force, $\bF^{net} \sim  k_sT_s \oint n_{out} J_S \bn dS $, vanishes because $n_{out}=n^\infty$ is a constant and the geometric integral $\oint J_S \bn dS$ vanishes for a body with a smooth shape. Thus, when $\lambda\delta^2/L\ll1$ the net force applied by swimmers on an arbitrary shaped body scales quadratically with $\lambda\delta^2/L$: $ \bF^{net}\sim O(\lambda\delta^2/L)^2 n^\infty k_sT_s L^2$. 

In Section 4 the analytical solutions are compared to particle-tracking simulations and direct numerical solutions to the Smoluchowski PDE. Both simulations and PDE solutions suggest a universal scaling not limited to $\lambda\delta^2/L\ll 1$ for the net force: ${F^{net}}/\left({n^\infty k_sT_s L}\right) = f\left({\lambda\delta^2}/{L}\right)$, where $f(x)$ is a dimensionless function.
The quadratic scaling $\bF^{net}\sim O(\lambda\delta^2/L)^2 n^\infty k_sT_s L^2$ when $\lambda\delta^2/L\ll 1$, or $f(x\to 0) \sim O(x^2)$, is verified by both simulations and PDE solutions. When $\lambda\delta^2/L\sim 1$, the effects of curvature still scale with this parameter, but show a linear dependence for the force, $\bF^{net}\sim O(\lambda\delta^2/L)n^\infty k_sT_s L^2$, or $f(x)\sim O(x)$ for $x\sim 1$. 

In Section 5 we discuss the effects of curvature for different ratios of the three lengths $\delta$, $\ell$ and $L$. When $\lambda\delta^2/L\ll 1$, the net force $\bF^{net}$ scales quadratically with $\lambda\delta^2/L$ due to the disappearance of curvature variations as explained above. When $\lambda\delta^2/L\sim 1$, i.e. $\ell \sim L$, the quadratic dependence is no longer valid because the expansion in curvature must be continued to a higher order than what we performed in Section 3 to include the effects of the variations of the principal curvatures. When this variation is considered, in the inner region a tangential flux may appear and due to the continuity of swimmer number density a net flux of swimmers towards the boundary may appear in the outer region, and therefore the outer solution $n_{out}$ is no longer a constant. As a result  a linear dependence of $\bF^{net}$ on $\lambda\delta^2/L$ may appear.

We then discuss the connection of this boundary-layer theory to a continuum mechanics point of view\cite{SwimForce2015} in an analogy to rarefied gas dynamics, and a formulation necessary to construct the general exact solution to the number density $n$ and polar order $\bmm$ fields with an arbitrary shaped body. We conclude this work with a brief discussion  including  hydrodynamics into this pure kinetic boundary-layer analysis.

In this work  we only consider a uniform number density and random orientation far from the body.  When these restrictions are relaxed,  as for example when there is a net directed swimming motion of the active particles, even a spherical body can experience a net force and therefore motion.\cite{Brady2017}

\section{Problem formulation}
\begin{figure}[h]
	\centering
	\includegraphics[width=\linewidth]{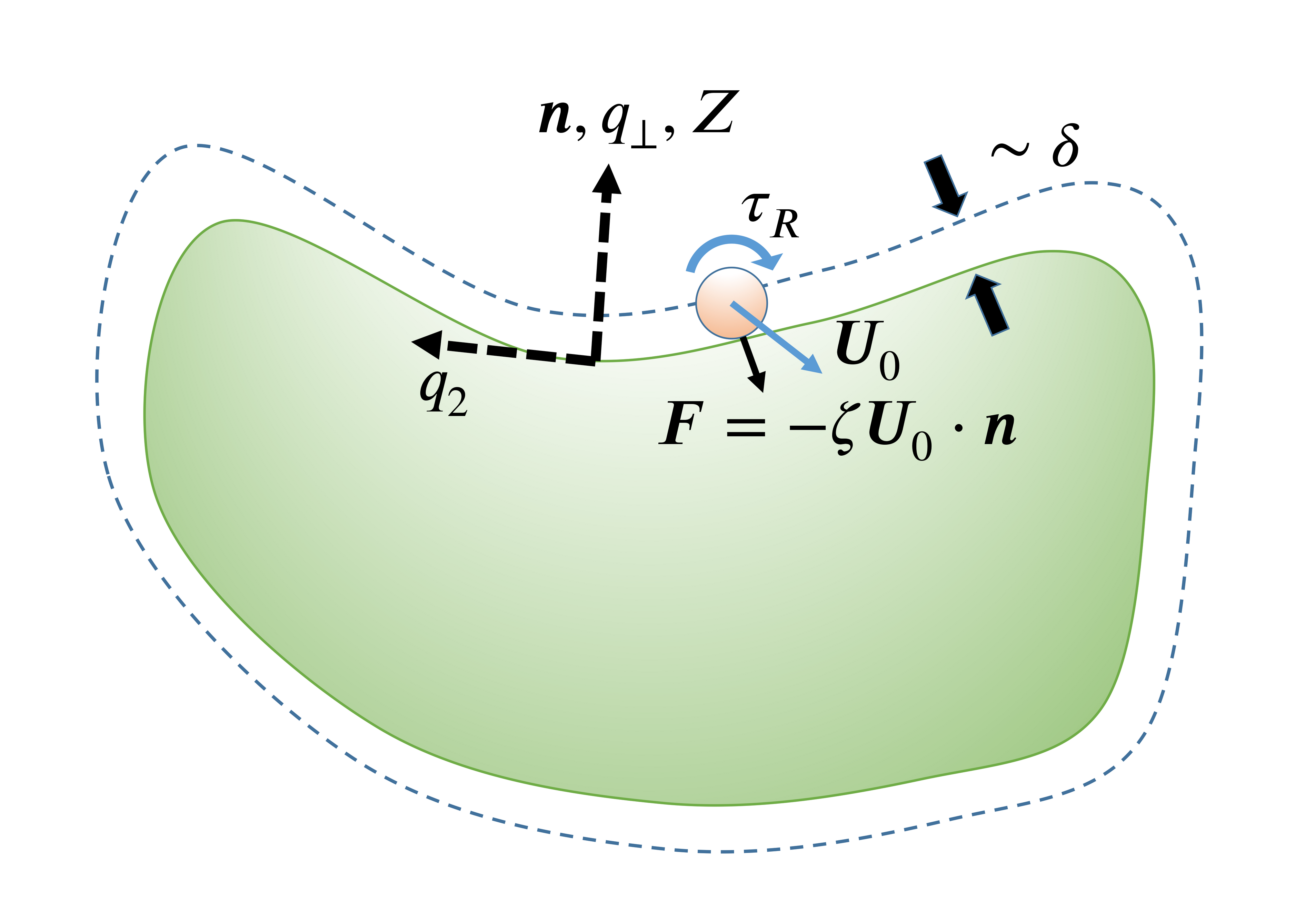}
	\caption{\label{fig:BLschematic} The formation of the boundary layer on the body surface. When a swimmer comes to the surface, it transmits a force of $\bF=-\zeta \bU_0\cdot\bn$ to the boundary. The swimmers form an accumulation boundary layer with thickness on the order of $\delta=\sqrt{D_T\tau_R}$. The inner solution is solved in the local coordinate system $q_\perp,q_2$, depending on the local body curvature. The definition of the local curvature is shown in Figure~\ref{fig:geocoordinate}.}
\end{figure}

The behavior of ABPs depends on (at least) three characteristic lengths:  (i) the macroscopic length scale $L$ of the body, (ii) the run length $\ell = U_0\tau_R$, and (iii) a microscopic length $\delta = \sqrt{D_T \tau_R}$, where $D_T$ is the translational diffusivity of the active particles (see Figure~\ref{fig:BLschematic}). When the reorientation process is due to rotary Brownian motion, the microscopic length is proportional to the active particle size, $\delta = \sqrt{4/3}a$ for spherical ABPs. For a typical swimming micro-organism or a synthetic Janus particle, $\tau_R\sim \SI{1}{\second}$ , $\delta \sim \SI{1}{\micro\meter}$, and $\ell\sim 1-\SI{10}{\micro\meter}$.

The probability density for finding an active particle at position $\bx$ with orientation $\bq$ at time $t$ relative to the macroscopic body  is governed by the Smoluchowski equation:
\begin{equation}
	\dpone{P(\bx,\bq,t)}{t} + \nabla \cdot \bj^T + \nabla_{R} \cdot \bj^R = 0\,.
		\label{eq:SmolCurve}
\end{equation}
In the dilute limit, which we consider here, the translational and rotational fluxes are are given by
\begin{eqnarray}
	\bj^T =   (U_0 \bq  - D_T \nabla \ln P)P\,, \\
	\bj^R = - D_R \nabla_{R} P\,,
	\label{eq:fluxes}
\end{eqnarray}
where $\nabla_R = \bq\times \nabla_{\bq}$ is  the orientational gradient operator.\cite{Brenner1972} For a spherical swimmer of radius $a$ in a Newtonian solvent of viscosity $\eta$, $\zeta=6\pi\eta a$, $D_T = k_BT/\zeta$, $D_R =1/\tau_R = k_BT/8\pi \eta a^3$ and $\delta = \sqrt{D_T/D_R}$. In this paper we develop a general theory, and thus allow both $D_T$ and $D_R$ to be arbitrary. 

At a boundary, ABPs cannot cross the surface and thus the normal component of the translational flux must vanish:
\begin{align}
	\bn\cdot\bj^T =0 \,.
\end{align}
Far from the body at infinity, we assume the swimmers are unperturbed with uniform number density $n^\infty$ with an unbiased orientation distribution.

The conservation equations for the zeroth and first moments of the Smoluchowski equation are:\cite{Saintillan2015}
\begin{align}
\frac{\partial n}{\partial t}  +   \nabla\cdot \bj_n = 0, \,\,\,& \bj_n  =  U_0\bmm - D_T \nabla n\,, \label{eq:jn} \\
\frac{\partial \bmm}{\partial t}  +   \nabla\cdot  \bj_m + 2D_R \bmm = 0,\,\,\,& \bj_m =    U_0\bQ  +  {\frac{1}{3}}U_0 n\, \bI - D_T \nabla\bmm\,, \label{eq:jm} 
\end{align}
where  $\bmm(\bx,t) = \int \bq P(\bx,\bq,t) \mathrm{d}\bq$ is the polar order field,  and $\bQ(\bx,t) = \int (\bq\bq- \frac{1}{3}\bI) P(\bx,\bq,t)\mathrm{d}\bq$ is the (zero-traced) nematic order field. The hierarchy can be continued to include an equation for $\bj_Q$, allowing spatial variation of the nematic order. It has been shown\cite{ForceBoundary2015} that including a nematic order does not significantly improve the boundary-layer solution. The wall introduces an asymmetry that is either towards or away from the wall, and therefore the polar order is  the most important moment. Including higher moments only slightly improves the solution in the limit of $\ell\gg\delta$ and $\ell\sim L$.

To understand the structure of the problem and the appearance of a screening length and boundary layer, we non-dimensionalize the equations with the macroscopic length scale $L$ and consider the steady state only:
\begin{align}
\hnabla\cdot\bm{j}_n &= 0\,,\label{eq:jnnonD}\\
\hnabla\cdot\bm{j}_{\bm{m}} + 2\bm{m}&= 0\label{eq:jmnonD}\,,
\end{align}
with fluxes:
\begin{align}\label{eq:jmnnonD}
\bm{j}_n &= \frac{\ell}{L} \bm{m} - \frac{\delta^2}{L^2}\hat{\nabla} n\,,\\
\bm{j}_{\bm{m}} &= \frac{\ell}{3L} n\bI -\frac{\delta^2}{L^2}\hat{\nabla} \bm{m}\,,
\end{align}
where  $\hat{\ }$ denotes non-dimensionalization with $L$.
The truncation $\bQ=0$ allows a simple mathematical manipulation: Setting $f=\hat{\nabla}\cdot\bmm$, taking the divergence of~(\ref{eq:jmnonD}) and eliminating $\hnabla^2n$ with~(\ref{eq:jnnonD}), we have
\begin{align}
\left(\hat{\nabla}^2 - \lambda^2 L^2 \right) f&= 0\,, \label{eq:mdivscreen}\\
\hat{\nabla}^2 n &= \frac{L\ell}{\delta^2}f\,,\label{eq:nlaplace}
\end{align}
where 
\begin{equation}
\lambda = \sqrt{2\left[1+\tfrac{1} {6}\left(\ell/\delta\right)^2\right]}/\delta
\label{eq:lambda}
\end{equation}
is the inverse screening length.\cite{ForceBoundary2015}

Equation~(\ref{eq:mdivscreen}) is a homogeneous Helmholtz equation, which has a `screened' -- exponentially decaying -- solution. Since we know in free space far from the body $n=n^\infty=const$,  $\bmm=0$, and $f^\infty=0$, and therefore $f$ must decay exponentially on the  scale $\lambda L$ away from the boundary.

Equation~(\ref{eq:nlaplace}) is an inhomogeneous Laplace equation, and the solution can be decomposed into a homogeneous general solution $n_H$ and an inhomogeneous particular solution $n_P$:
\begin{align}
	n&=n_H+n_P\,,\\
	\hnabla^2 n_H &=0,\quad\hnabla^2n_P = \frac{L\ell}{\delta^2}f\,.
\end{align}
Therefore, we know that $n_H$ decays algebraically, e.g. as $1/r$, governed by  Laplace's equation, and $n_P$ decays at the same rate as $f\  (=\hnabla\cdot\bmm)$.

Once we know $n=n_H+n_P$, we can substitute it back into the equation to solve for $\bmm$:
\begin{align}
 \frac{\delta^2}{L} \hnabla^2\bmm -2\bmm = \frac{\ell}{3L} \nabla n\,.
\end{align}
Again, due to the structure of this inhomogeneous Helmholtz equation, $\bmm$ can be decomposed into a homogeneous general solution and a particular solution depending on $\nabla n$; $\bmm=\bmm_H+\bmm_P$, and
\begin{align}
\frac{\delta^2}{L^2} \hnabla^2\bmm_P -2\bmm_P &= \frac{\ell}{3L} \hnabla n\,, \\
\frac{\delta^2}{L^2} \hnabla^2\bmm_H -2\bmm_H &= 0\,.
\end{align}

With some mathematical construction, we can explicitly calculate the particular solutions $n_P$ and $\bmm_P$, which are:
\begin{align}\label{eq:mnparticular}
	n_P &=  \frac{\ell}{\delta^2\lambda^2 L} \hnabla\cdot\bmm\,, \\
	\bmm_P &=  \frac{1}{\lambda^2L^2} \hnabla(\hnabla\cdot\bmm) - \frac{\ell}{6L} \hnabla n_H\,,
\end{align}
The particular solution $\bmm_P$ contains the long-ranged part depending on $\hnabla n_H$\, .

To summarize, the general structure of the problem is:
\begin{itemize}
\item $f=\hnabla\cdot\bmm$ exponentially decays as $\exp{\left(-\lambda L\hr\right)}$\,.
\item $n_H$ is long ranged as $n^\infty + O(1/\hr)$\,.
\item $n_P\sim f$ and decays exponentially as $\exp{\left(-\lambda L\hr\right)}$\,.
\item $\bmm_H$ decays exponentially as $\exp{\left(-L\hr/\delta\right)}$\,.
\item $\bmm_P$ contains both an exponential $\exp{\left(-\lambda L\hr\right)}$ and a long ranged, $O(1/\hr^2)$, component.
\end{itemize}

The general structure of the solution shows that as long as $\delta \ll L$ there will be a region near the body with rapid decay of part of the concentration and polar order fields.
The solutions can thus be split into an (exponential) inner region and an $O(1/\hr)$ outer region. Only $n_H$ (and the corresponding component in $\bmm_P$) extends to the outer region and is governed by Laplace's equation~(\ref{eq:nlaplace}). Inside the boundary layer, due to the separation of scales, the outer solution $n_H$ can be considered as linear or even a constant.
The inner region is attached to the body surface and can be considered as an 1D (curved) boundary layer. After the boundary layer is solved, it serves as the boundary condition for  Laplace's equation in the outer region. The final solution can then be determined by matching the flux between those two regions, which is a standard boundary-layer approach.

From the definition of $\lambda$ in (\ref{eq:lambda}), when $\ell\sim\delta$, we have $\lambda \sim 1/\delta$ and $n_P$ and $\bmm_H$ decay at comparable rates. When $\ell\gg\delta$, we have $\lambda \sim \ell/\delta^2 \gg 1/\delta$, and $n_P$ decays much faster than $\bmm_H$. In this case, the inner region of thickness $\delta$ further splits into two regions of different scales, and forms `a boundary layer inside a boundary layer.' Our analysis in the following section, however, is general and applies to both cases. The `boundary layer' includes the exponential part of the solution, and the portion of the slowly decaying part of the solution that is  located within the thin layer. As we shall see later, the slowly decaying part inside the boundary layer is approximated as a linear function. In sum, we shall solve the boundary layer structure with the assumption that $\delta \ll L$ and $\ell \ll L$, but we do not specify the relation between $\delta$ and $\ell$. Our analytical solution requires both $\delta$ and $\ell$ to be much smaller than $L$ because of the curvature expansion we develop  inside the boundary layer is only valid in this limit. The regime $\ell\sim L$ requires a mathematically formidable high order expansion in curvature, which will become clear in the next section (cf.~(\ref{eq:divfinner})).

\section{Analytical solution}
Inside the boundary layer, the solution depends on the local geometry only. We  split the solution into an inner region, $\left(n_{in}, \bmm_{in}\right)$, and an outer region, $\left(n_{out}, \bmm_{out}\right)$. All lengths are scaled with the macroscopic length  $L$. As discussed in the last section, the outer region is governed by Laplace's equation:
\begin{align}
	\hnabla^2 n_{out} =0\,,
\end{align}
while in the inner region we solve the full equations, starting with~(\ref{eq:mdivscreen}). For a flat boundary with vanishing curvature ($\kappa_1=\kappa_2=0$ in the notation below), the boundary layer has been solved.\cite{ForceBoundary2015} Inside the boundary layer, we rescale the coordinate normal to the flat plate, $q_\perp$, as $Z=q_\perp/\left(\delta/L\right) = q_\perp / \epsilon$, where $\epsilon=\delta/L\to 0$, and the solution is:
\begin{subequations}
		\label{eq:flatinner}
	\begin{align}
	\frac{f_0}{n^\infty} &= \frac{\ell L}{\delta^2} \frac{6\delta^2 + \ell^2}{18 \delta^2} e^{-\lambda \delta Z}\,, \\
	\frac{n_0}{n^\infty} &= 1 + \frac{\ell^2}{6\delta^2} e^{-\lambda \delta Z}\,,\\
	\frac{m_{\perp,0}}{n^\infty} &= -\frac{\ell\lambda}{6} e^{-\lambda \delta Z},\quad m_{\parallel,0}=0\,,
	\end{align}
\end{subequations}
where the subscript $0$ means zero curvature of the wall.
For a flat wall geometry, the outer solution is simply $n_{out}=n^\infty=const$ and the pressure exerted on the flat wall by the active particles is
\begin{align}
\Pi_{wall}&=n_{wall} \zeta D_T  = \left(1 + \frac{\ell^2}{6\delta^2}\right)n^\infty \zeta D_T\, , \nonumber \\ 
&= \zeta n^\infty\left(D_T+D^{swim}\right) = n^\infty\left(k_BT + k_sT_s\right)\,.
\end{align} 
For a body of any shape, to zeroth order in the body curvature, i.e. to leading order in $\delta/L$ (and in $\ell/L$ since there is no macroscopic length for the 1D flat plate; $L \rightarrow \infty$), the above concentration profile holds and the pressure on the body is the same everywhere on its surface. The net force on any body given by this homogeneous pressure is zero, just as a macroscopic body submersed in the atmosphere does not experience a net force from the homogeneous isotropic atmospheric pressure. For a classical ideal gas, the pressure on a boundary would deviate from the ideal gas limit when the mean free path is comparable to the macroscopic body length. For ABPs, the pressure is applied to the body surface through the formation of the kinetic boundary layer, and there are two length scales, $\delta$ and $\ell$, related to this boundary layer. So the pressure would deviate from the isotropic swim pressure, $n^\infty\left(k_BT + k_sT_s\right)$, when either $\delta$ or $\ell$ is comparable to the body's macroscopic size. With the boundary-layer analysis we find the correction to the swim pressure on the wall appears at the order $\lambda\delta^2/L$.

\begin{figure}[h]
	\centering
	\includegraphics[width=\linewidth]{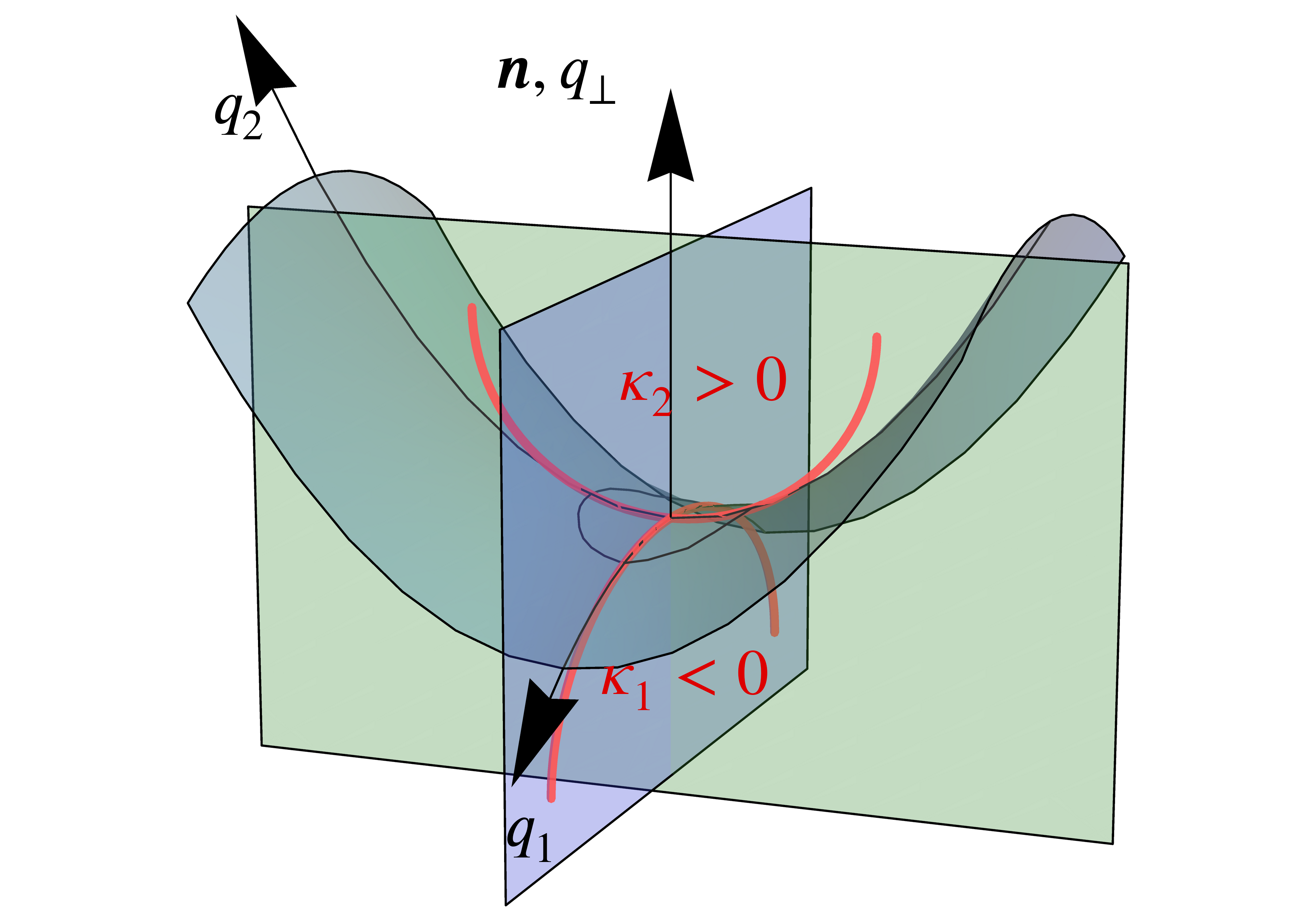}
	\caption{\label{fig:geocoordinate} The local representation of an arbitrary curved surface. The two principal curvature directions are located in the two perpendicular planes. The local curvilinear coordinate system $q_1,q_2,q_\perp$ is built on the surface, where locally it is orthogonal. $q_\perp$ aligns with the normal vector $\bn$ pointing toward the outside of the shape at that point. The sign of principal curvatures $\kappa_1$ and $\kappa_2$ follows the convention shown here.}
\end{figure}

The boundary-layer thickness is governed by the microscopic length $\delta=\sqrt{D_T\tau_R}$, and therefore we must improve upon the zero-curvature solution in order to find the conditions for a net force on a body.  We build a local curvilinear coordinate system as shown in Figure~\ref{fig:geocoordinate}. The arbitrary curved surface is represented by a second order curvature (mathematically, the second fundamental form). Coordinate axes $q_1,q_2$ are attached to the curvature surface along the two principal curvature directions and $q_\perp$ is aligned with the normal vector $\bn$. We can assume that locally $q_1,q_2,q_\perp$ are orthogonal.

In the curvilinear coordinate system, the Cartesian nabla operator $\hnabla$ in~(\ref{eq:mdivscreen}) is replaced by the curvilinear nabla operator $\nablas$ (cf. the Appendix). Here the $\hat{\ }$ sign also means that these operators are non-dimensionalized by the macroscopic length $L$, which is the same for the curvature $\kappa$: $\hkappa=L\kappa$. The details of this curvilinear nabla operator can be found in the Appendix, and to  leading order, $O(\hkappa\delta/L)$, we need only consider a constant $J_S=2H$, where $H=\frac{1}{2}(\hkappa_1+\hkappa_2)$ is the mean (non-dimensionalized) curvature. Also, the first order effects give only a correction in the normal direction; the  solution in the tangential direction appears at second order.  

Here we consider a smooth body and assume the curvature $\hkappa \sim O(1)$ everywhere. If there is some non-smoothness, for example, a sharp tip on the body, the curvature $\hkappa_{tip} \to -\infty$, and the boundary-layer assumption is no longer valid there, so a leading order curvature solution is not sufficient. Physically, swimmers with orientation $\bq\neq - \bn$ have a tangential swim velocity and are able to easily leave the sharp tip; this is also true for a passive particle with significant translational diffusivity $D_T$. Therefore, the kinetic accumulation on a sharp tip is very weak, and the pressure there significantly decreases with more negative curvature.\cite{ForceBoundary2015} However this local curvature argument is only applicable to the inner boundary layer solution, which will become clear after we reach the solution~(\ref{eq:nwallcurvature}) and~(\ref{eq:piswimjs}). We will revisit this issue in Section 4 with the example in Figure~\ref{fig:denmap}.

\subsection{Inner solution}
In the boundary layer, we denote the number density at the top of the boundary layer as $\ntop$, which is the local value of the outer solution and is not necessarily a constant as is $n^\infty$. The governing equation for $f$ in curvilinear coordinates is
\begin{align}\label{eq:divminner}
	\left(\nablas^2 - \lambda^2 L^2\right)f&=0\,,
\end{align}	
which becomes
\begin{align}
	\left( -J_S\dpone{ }{q_\perp} + \dptwo{ }{q_\perp} - \lambda^2L^2\right)f&=0\,,
	\label{eq:divfinner}
\end{align} 
where $J_S\sim O(\hkappa) \sim O(1)$. Inside the boundary layer we rescale the perpendicular coordinate as $Z=q_\perp/\epsilon$, where $\epsilon\sim\delta/L\ll 1$ and $Z\sim O(1)$. Thus, (\ref{eq:divfinner}) becomes
\begin{align}
	\left( -J_S {\epsilon} \dpone{ }{Z} +  \dptwo{ }{Z} - \lambda^2\delta^2 \right)f = 0\,.
\end{align}
Note that here by taking the leading order curvature expansion of the operator $\nablas$, we assumed that $\ell\ll L$. Otherwise if $\ell\sim L$, $\lambda^2L^2$ scales as $(L/\delta)^4$ and the leading order curvature expansion is not sufficient. We will see that this restriction $\ell \ll L$ naturally appears in the parameter $\lambda\delta^2/L$ in the final result of this boundary-layer analysis.
  
The leading order effects of the curvature are captured by the asymptotic expansion:
\begin{align}
	f&=f_0 + \epsilon f_1 + ...\\
	n&=n_0 + \epsilon n_1 + ...\\
	m_\perp &= m_{\perp,0} + \epsilon m_{\perp,1} + ...\, ,
\end{align}
where the leading order $f_0$, $n_0$ and $m_{\perp,0}$ are just the flat surface solution~(\ref{eq:flatinner}). We show later that $\bmm_\parallel$ remains zero at this order.

From the governing equations we see that if $\ell/\delta\sim O(1)$, then $f_0/\ntop\sim  L/\delta$, $n_0/\ntop\sim O(1)$, $m_{\perp,0}/\ntop\sim O(1) $. If, however, $\ell\gg\delta$, then $f_0/\ntop\sim \ell^3 L/\delta^4$, $n_0/\ntop\sim\ell^2/\delta^2$, and $m_{\perp,0}/\ntop\sim \ell^2/\delta^2 $, which are not on the same order due to the different prefactors in (\ref{eq:flatinner}). Thus,  we need to be  careful when going to the next order to maintain these scalings. 

At first order
\begin{align}
\left( \dptwo{ f_1 }{Z} - \lambda^2 \delta^2 f_1 \right)  &=  J_S \dpone{f_0}{Z}\, ,
\end{align}
with solution
\begin{align}
\label{eq:finner}
\frac{f_1}{\ntop} & =  C_1 e^{\delta  \lambda  Z}+ C_2 e^{-\delta  \lambda  Z }\nonumber \\
&+\frac{1}{12} \ell J_S \lambda ^2 L Z e^{-\delta  \lambda  Z } +\frac{\ell J_s \lambda  L }{24 \delta }e^{-\delta  \lambda  Z }\, ,
\end{align}
where $C_1$ and $C_2$ are constants to be determined. As discussed in the last section, $f$ is screened so all components of $f_1$ are exponential. We require $C_1=0$ because an exponentially growing term is not possible. 

With $f_1$ we can then solve for the first curvature correction $n_1$:
\begin{align}
	\dptwo{n_1}{Z} = \frac{\ell}{L}f_1 + J_S \dpone{n_0}{Z}\,,
\end{align}
with solution:
\begin{align}\label{eq:ninner}
	\frac{n_1}{\ntop}&=\frac{C_1 \ell e^{\delta  \lambda  Z}}{\delta ^2 \lambda ^2 L}+\frac{{C_2} \ell e^{-\delta  \lambda  Z }}{\delta ^2 \lambda ^2 L}+{C_3}+{C_4} Z \nonumber \\
	&+\frac{\ell^2 J_S e^{-\delta  \lambda  Z }}{24 \delta ^3 \lambda }+\frac{\ell^2 J_S Z e^{-\delta  \lambda  Z }}{12 \delta ^2}\,.
\end{align}

The last step is to solve for $m_\perp$:
\begin{align}
	-J_S \dpone{m_{\perp,0}}{Z} + \dptwo{m_{\perp,1}}{Z} - 2 m_{\perp,1} = \frac{\ell}{3L}\frac{L}{\delta} \dpone{n_1}{Z}\, ,
\end{align}
with solution
\begin{align}\label{eq:mperpinner}
	\frac{m_{\perp,1}}{\ntop} &=\frac{{C_1} e^{\delta  \lambda  Z}}{\lambda  L}-\frac{C_2 e^{-\delta  \lambda  Z }}{\lambda  L}-\frac{1}{12} \ell J_S \lambda  Z e^{-\delta  \lambda  Z }\nonumber \\
	&+\frac{\ell J_S e^{-\delta  \lambda  Z }}{24 \delta } -\frac{\ell}{6\delta}C_4\,.
\end{align}
Note that the solution for $m_{\perp,1}$ agrees with the requirement that $f=\nablas\cdot\bmm$.

The parallel component of the polar order $\bmm_{\parallel}$ remains zero at  first order,  since the curvature expansion of $\nablas$ in~(\ref{eq:divfinner}) only involves the perpendicular direction and  the zero-th order $\bmm_{\parallel}$ is zero. Physically this is because to leading order in curvature, the variation of curvature is ignored and no tangential flux can appear.

The structure of $f_1,n_1$ and $m_{\perp,1}$ follows the separation of scales as discussed in the last section. Particularly, the $C_3+C_4 Z$ part in $n_1$ represents the long-range solution $n_H$. The $-\ell C_4/(6\delta)$ part in $m_{\perp,1}$ is the long-ranged part, $-\ell \hnabla n_H/(6\delta)$, in $\bmm_P$. Inside the thin boundary layer, the variation of $n_H$ is slow, and is simplified to a linear function of $Z$ in the first order solution for $n_1$.

\subsection{Satisfying the boundary condition \&  flux}
First, the exponentially growing part $C_1$ must be zero. On the surface $Z=0$, the non-penetrating boundary condition is:
\begin{align}
	\bj_n\cdot\bn_Z &= \ell m_{\perp} - \frac{\delta^2}{L} \frac{L}{\delta} \dpone{n}{Z} =0\,, \\
	\bj_{\bmm} \cdot\bn_Z &= \dfrac{n}{3}\ell  - \frac{\delta^2}{L} \frac{L}{\delta}\dpone{m_{\perp}}{Z} =0\,.
\end{align}

\subsubsection{The boundary condition on $\bj_n$:}
\begin{align}\label{eq:jninnerzero}
	\bj_n\cdot\bn_Z &= \ell \left(m_{\perp,0}+\frac{\delta}{L}m_{\perp,1}\right) - \frac{\delta^2}{L}\frac{L}{\delta}\dpone{}{Z} \left( n_0 + \frac{\delta}{L}n_1 \right) \nonumber, \\
&= -\frac{C_4 \delta ^2}{L}\ntop\,,
\end{align}
which is a constant. To satisfy the boundary condition $\bj_n\cdot\bn_Z=0$, $C_4$ must be zero. The perpendicular component of the flux $j_{n,\perp}=\bj_n\cdot\bn_Z=0$ throughout the boundary layer. Therefore by the continuity of flux from the outer to the inner solution, this zero-flux boundary condition is also the boundary condition for the outer problem $\hnabla^2 n_{out}=0$.

\subsubsection{The boundary condition on $\bj_m$:}
By setting $\bj_m\cdot\bn_Z=0$ at $Z=0$ in~(\ref{eq:mperpinner}), we find:
\begin{align}
C_2 = C_3 \frac{ \ell L \left(6 \delta ^2+\ell^2\right)}{18 \delta ^4}+\frac{\ell J_S L \left(9 \delta ^2+2 \ell^2\right) \sqrt{\frac{\ell^2}{\delta ^2}+6}}{72 \sqrt{3} \delta ^4}\,,
\end{align}
where $C_3$ is found by matching to the outer solution for $n_1$: 
\begin{align}
\frac{n}{n_{top}} \to \frac{1}{n_{top}} \left[ n_0(Z\to\infty) + \frac{\delta}{L} n_1(Z\to\infty) \right] = 1+\frac{\delta}{L}C_3 \,.
\end{align}

\subsubsection{Divergence of the translational flux $\nabla\cdot\bj_n$:}
This is not a boundary condition, but we demonstrate that at steady state it is zero, as required by the governing equation~(\ref{eq:jn}).
The two leading orders to the flux $\bj_n$ are:
\begin{align}
	\nablas\cdot\bj_n &= \ell \frac{L}{\delta} \dpone{m_{\perp,0}}{Z} - L \dptwo{n_0}{Z} \nonumber \\
	& + \ell\left( -J_S m_{\perp,0} + \dpone{m_{\perp,1}}{Z}\right) - \delta \left( \dptwo{n_1}{Z} - J_S \dpone{n_0}{Z} \right)\,,
\end{align}
and we have that $m_{\perp,0} = \dfrac{\delta}{\ell} \dpone{n_0}{Z}$;
therefore,
\begin{align}
	\nablas\cdot\bj_n = \ell\left( \dpone{m_{\perp,1}}{Z}\right) - \delta \left( \dptwo{n_1}{Z} \right)\,.
\end{align}
From the solutions (\ref{eq:ninner}) and (\ref{eq:mperpinner}), $\nablas\cdot\bj_n=0$ everywhere inside the boundary layer, as required by the governing equations.

\subsection{The continuity of $j_{n,\perp}$ and the boundary condition on $n_{out}$.}
As discussed in the previous section, the solution for $n$ can be decomposed into a homogeneous,  $n_H$, and a particular,  $n_P$, solution. From the  boundary-layer perspective, $n_H$ obeys  Laplace's equation and is  the outer solution $n_{out}$. Thus, in the outer region,
\begin{align}
	\nabla^2 n_{out} &= 0\,,\\
	n_{out}(\hat{r}\to\infty)&= n^\infty\,,\\
	\text{boundary condition: }j_{n,\perp}&=\bj_n\cdot\bn=0\,.
\end{align}
According to~(\ref{eq:jmnnonD}) and~(\ref{eq:mnparticular}), in the outer region $\bmm=-\ell \hnabla n_H/(6 L)$, and therefore:
\begin{align}\label{eq:jnouter}
	\bj_n &= \frac{\ell}{L} \bmm_P - \frac{\delta^2}{L^2}\hnabla n_{out} = - \frac{\ell}{L} \frac{\ell}{6 L} \hnabla n_{out} - \frac{\delta^2}{L^2}\hnabla n_{out}, \nonumber \\
	&\sim  - \left(D^{swim}+D_T\right)\nabla n_{out} =0\,,
\end{align}
which then gives the boundary condition on the outer region $\hnabla n_{out} \cdot\bn=0$. 
By the uniqueness of the Laplace's equation subject to Neumann boundary condition, $n_{out}$ is simply:
\begin{align}
	n_{out} =const= n^\infty\,.
\end{align}
The outer solution is a constant everywhere, and so at the top of the boundary layer $\ntop=n^\infty$ and $C_3=0$. The inner solutions are given in (\ref{eq:ninner}) and (\ref{eq:mperpinner}), which decay exponentially away from the surface and scale as $n^\infty$.

\section{Results}
\subsection{Analytical results}
With the solution $\ntop=n^\infty$ and~(\ref{eq:ninner}) and~(\ref{eq:mperpinner}), we can calculate the swim pressure exerted by the kinetic boundary layer everywhere on an arbitrary shaped body with $\Pi_{wall}=\zeta D_T n_{wall}$:
\begin{align}\label{eq:nwallcurvature}
{n_{wall}} &= n_0(Z=0) + \frac{\delta}{L}n_1(Z=0),  \nonumber \\
&= n^\infty \left[ 1+ \frac{\ell^2}{6\delta^2} + \frac{\ell^2 \lambda }{12 L } J_S \right] \,.
\end{align}
Thus to  first order in the curvature $J_S$ we have:
\begin{align}\label{eq:piswimjs}
\Pi_{wall} &= n^\infty \zeta D_T + n^\infty \zeta D^{swim}\left(1+\frac{\lambda\delta^2}{2L} J_S\right)\,, \nonumber \\
&=n^\infty k_BT + n^\infty k_sT_s\left(1+\frac{\lambda\delta^2}{2L}J_S\right)\,.
\end{align}
For passive Brownian particles $D^{swim}=0$ and $\Pi_{wall}$ is not affected by the curvature $J_S$ as must be the case. For swimmers, the swim pressure is affected by the curvature and scales as $J_S\lambda \delta^2/(2L)$.

In the limit of fast swimmers $\ell\gg\delta$, the inverse screening length $\lambda\to {\ell}/({\sqrt{3}\delta^2})$, and: 
\begin{align}
\frac{n_{wall}}{n^\infty} \to 1+ \frac{\ell^2}{6\delta^2}\left[ 1+ \frac{1}{2\sqrt{3}} \frac{\ell}{L} \left(\hkappa_1+\hkappa_2\right) \right]\,.
\end{align}
In this regime the swim pressure becomes:
\begin{align}\label{eq:pi3dellL}
\Pi_{wall} \to n^\infty k_BT + n^\infty k_sT_s\left[ 1+ \frac{1}{2\sqrt{3}} \frac{\ell}{L} \left(\hkappa_1+\hkappa_2\right) \right]\,,
\end{align}
where $\hkappa_1$ and $\hkappa_2$ are the two signed non-dimensional principal curvatures.
This result agrees with our previous study of swimmers outside a sphere.\cite{ForceBoundary2015} The signed curvatures $\hkappa_1$ and $\hkappa_2$ follow the sign convention in Figure~\ref{fig:geocoordinate}, and therefore the pressure on the boundary {\em decreases} as the sphere decreases in size.
It is also consistent with the behavior in the singular limit of no translational Brownian motion $D_T\to0$,\cite{Smallenburg2015,FilyConfinedSwimmers2014} where the pressure depends only on the ratio of run length to the macroscopic length and curvature $\hkappa\ell/L$. 

The above analytical solutions to leading order apply when the microscopic length $\delta$ and the run length $\ell$ are both much smaller than the macroscopic body length $L$, and therefore they are accurate only to the $O(\delta/L)$ and $O(\ell/L)$. 
The derivation shows that the correct small parameter that encompasses both of these  limits is $\lambda \delta^2/L$:
\begin{subequations} \label{eq:limitscaling}
\begin{align}
\text{when } &\ell\sim\delta: && \lambda\delta^2/L\sim \delta/L\,, \\
\text{when } &\ell\gg\delta: && \lambda\delta^2/L \sim \ell/L\,,
\end{align}
\end{subequations}
Therefore our analysis applies to leading order in $\lambda\delta^2/L$ as shown in~(\ref{eq:nwallcurvature}), which we verify below by comparison to simulation data.

Equation~(\ref{eq:piswimjs}) gives the pressure distribution everywhere on an arbitrary shaped body. With a surface integration, we  get the net force and torque on the body. Clearly, integration of the constant part of pressure, $n^\infty \left( k_BT + k_sT_s\right)$, does not give a net force or torque, and thus to leading order:
\begin{subequations}\label{eq:integral3D}
\begin{align}
\oint \Pi^{swim} dS &= n^\infty k_s T_s L^2 \oint \left(1+\frac{\lambda\delta^2}{2L} J_S\right) d\hat{S}\,, \\
\bF^{net} &= n^\infty k_s T_s L^2 \oint -\frac{\lambda\delta^2}{2L} J_S \bn d\hat{S}\,, \\
\bL^{net} &= n^\infty k_s T_s L^2 \oint -\frac{\lambda\delta^2}{2L} J_S \br\times \bn d\hat{S}\,.
\end{align}
\end{subequations}
Here, $\oint \Pi^{swim} dS$ is a scalar `total' integration of the swim pressure on the body, which is of no dynamic importance, but it can be easily measured from particle-tracking Brownian dynamics simulations and will help to verify the boundary-layer solution. The net force and torque, $\bF^{net}$ and $\bL^{net}$, arise solely from the asymmetric body shape.

Equation~(\ref{eq:integral3D}) involves the pure geometric integral of $J_S\bn$ over the body surface. By definition:
\begin{align}
J_S = 2H=(\hkappa_1+\hkappa_2)=-\hnabla\cdot\bn\,.
\end{align}
And it is well-known that for a smooth closed simply-connected surface:
\begin{align}
	\oint (\nabla\cdot\bn)\bn dS =0\,.
\end{align}
Therefore, we conclude that to first order in $O(\lambda\delta^2/L)$:
\begin{align}
	\bF^{net} = 0\,,
\end{align} 
For a smooth body of arbitrary shape a net force can only appear at $O(\lambda\delta^2/L)^2$.

Since to first order $\bF^{net}=0$, this means that the torque is a force couple that does not depend on the choice of the origin of the torque. If we shift the torque moment center by $\br_0$ in~(\ref{eq:integral3D}c), then:
\begin{align}
	\bL_0^{net} &= n^\infty k_s T_s L^2 \oint -\frac{\lambda\delta^2}{2L} J_S (\br+\br_0)\times \bn d\hat{S}\nonumber\,, \\
	&= \br_0\times \bF^{net} + \bL^{net} = \bL^{net}\,.
\end{align}
This equality holds at leading order in $\lambda\delta^2/L$. However, we are not able to get a general result as in the case for $\bF^{net}$, because we are not aware of a general mathematical answer for the geometric integral $\oint J_S \br\times\bn d\hat{S}$. We suspect, however, that the torque is also zero to leading order. 

\subsection{Verification}
To verify our analytical solution, especially the surprising result that to leading order the net force is zero, we performed Brownian dynamics simulations with ABPs and also numerically solved the PDEs, (\ref{eq:jnnonD}) and~(\ref{eq:jmnonD}), with a Finite Element solver. The simulations require as many as $40,000$ particles and times as long as $3000\tau_R$ to capture the very weak signal of $\bF^{net}$ in the large amount of Brownian noise. Also the Finite Element solver is expensive to solve due to the very thin boundary layer on the small scale $\delta$. Thus, we performed simulations and numerical solutions in a 2D geometry for ABPs with 2D in-plane rotations. 

In the 2D case, the boundary-layer analysis gives the same results only with a quantitative change due to the 2D reorientation:
\begin{align}
\frac{n_{wall}^{2D}}{n^\infty} = 1+ \frac{\ell^2}{2\delta^2} + \frac{\ell^2 \lambda' }{ L } J_S'\,,
\end{align}
where $J_S'=\hkappa$ is simply the (non-dimensional) curvature of the 2D curved boundary. For a curved boundary in 2D, there is only one curvature and there is no need to define a `mean curvature' $H$.
The 2D inverse screening length now becomes\cite{ForceBoundary2015}  $\lambda'=\sqrt{\left(1+\frac{1} {2}\left(\ell/\delta\right)^2\right)}/\delta$, and to leading order in $\lambda'\delta^2/L$ the 2D swim pressure (or tension) is:
\begin{align}\label{eq:piswimjs2D}
\Pi_{wall}^{2D} &= n^\infty \zeta D_T + n^\infty \zeta D^{swim}\left(1+\frac{\lambda'\delta^2}{L}J_S'\right), \nonumber \\
&=n^\infty k_BT + n^\infty k_sT_s'\left(1+\frac{\lambda'\delta^2}{L}J_S'\right)\,,
\end{align}
where $k_sT_s'=\zeta U_0^2/(2D_R)$. Also in the limit of $\ell\gg\delta$, similar to the 3D case~(\ref{eq:pi3dellL}), the perturbation to the swim pressure scales as $\ell/L$:
\begin{align}\label{eq:pi2dellL}
\Pi_{wall}^{2D} \to n^\infty k_BT + n^\infty k_sT_s'\left(1+\frac{\ell}{\sqrt{2} L}J_S'\right)\,.
\end{align}
This is consistent with the previous solution for active swimmers inside and outside a circle.\cite{ForceBoundary2015}

The net force and torque become:
\begin{subequations}\label{eq:integral2D}
	\begin{align}
	\oint \Pi^{swim} dL &= n^\infty k_s T_s' L \oint \left(1+\frac{\lambda'\delta^2}{L} J_S'\right) d\hat{L}\,, \\
	\bF^{net} &= n^\infty k_s T_s' L \oint -\frac{\lambda'\delta^2}{2L} J_S' \bn d\hat{L}\,, \\
	\bL^{net} &= n^\infty k_s T_s' L \oint -\frac{\lambda'\delta^2}{2L} J_S' \br\times \bn d\hat{L}\,.
	\end{align}
\end{subequations}
As before, $\bF^{net}=0$ to first order in the curvature. 

Actually, in 2D we can further simplify~(\ref{eq:integral2D}a) as a special case of Gauss-Bonnet theorem for a smooth 2D simple curve:
\begin{align}
	\oint \kappa d\hat{L} = -2\pi \,,
\end{align}
where the negative sign appears due to our sign convention as illustrated in Figure~\ref{fig:geocoordinate}.
Mathematically speaking, the `total curvature' is $-2\pi$ for a closed immersed plane curve. Hence,
\begin{align}\label{eq:ptot2Dcircum}
\oint \Pi^{swim} dL &= n^\infty k_s T_s'C \left(1 - 2\pi \frac{\lambda'\delta^2 }{C}\right)\,,
\end{align}
where $C$ is the circumference of the 2D shape  (on the scale of the macroscopic length $L$). 

The 2D Brownian dynamics simulation is done with the discretized Langevin equation of ABPs, with its orientation $\bq=	(\cos\theta,\sin\theta) $:
\begin{align}
	\Delta \bX &= U_0\bq\Delta t + \Delta X^B + \bF^C/\zeta \Delta t, \\
\Delta\theta &= \Delta \theta^B\,,
\end{align}
and $\AVE{\Delta X^B}=0$, $\AVE{\Delta X^B\Delta X^B}=2D_T\Delta t$,  $\AVE{\Delta \theta^B}=0$, and $\AVE{\Delta \theta^B\Delta \theta^B}=2D_R\Delta t$.
Particle-particle collision is ignored as that effect is not included in the dilute kinetic boundary-layer solution. The ABPs have radii $a$, and the swimmer-body collision force $\bF^C$ is applied through an excluded volume interaction at the contact line calculated with the potential-free algorithm.\cite{Foss2000,ForceBoundary2015} Due to the finite ABP radius $a$, the effective body shape is the original body shape plus a excluded volume layer of size $a$. The effective body shape is used in all data we present.

\begin{figure}
	\centering
	\includegraphics[width=\linewidth]{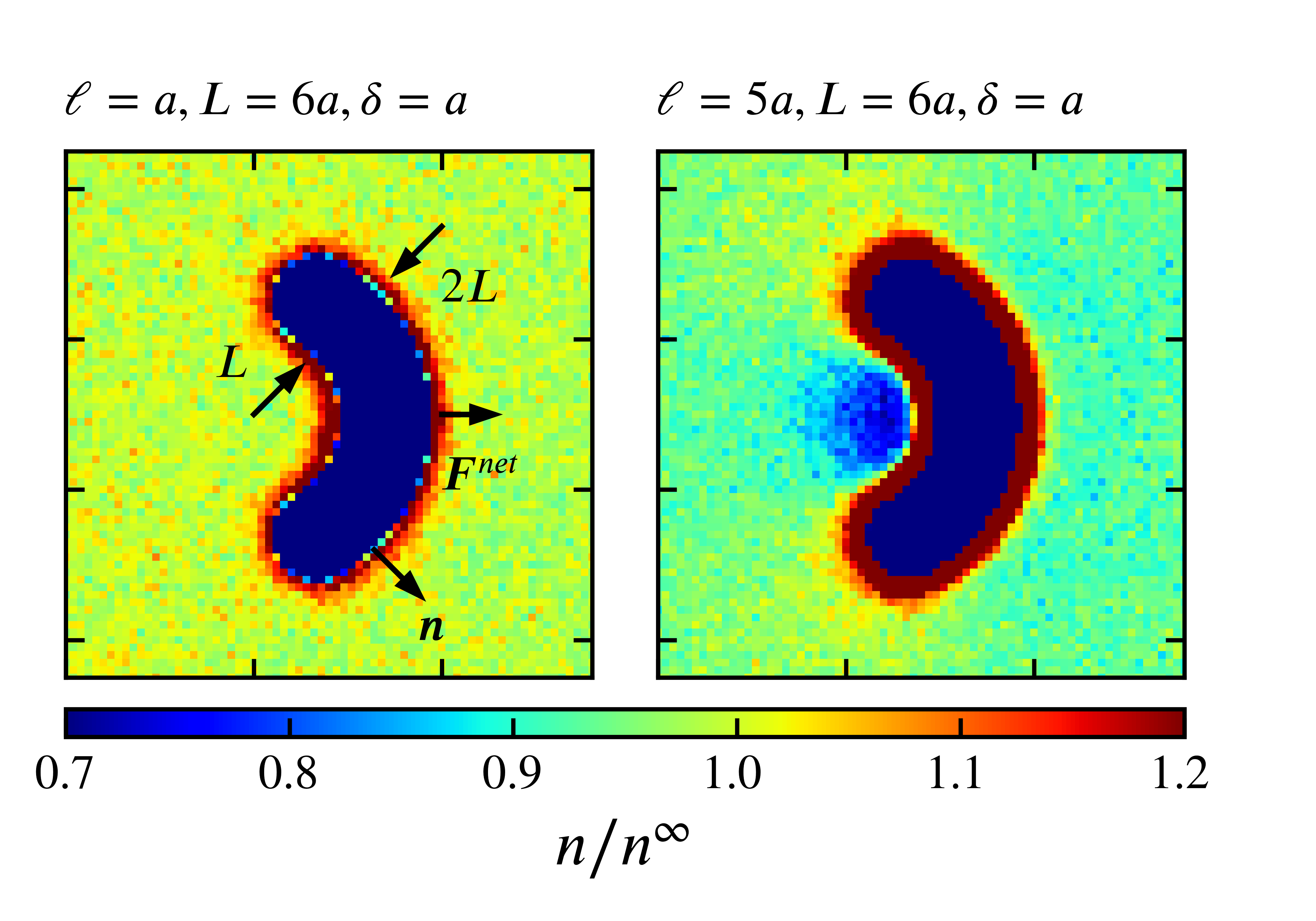}
	\caption{The number density $n/n^{\infty}$ for two cases. The length scale $L$ is set as the curvature radius of the concave part of the shape, and the convex part has a curvature radius $2L$. The concave and convex arcs are connected by two semicircular arcs to ensure that the shape is smooth everywhere. On the left, $\lambda'\delta^2/L={1}/{(2 \sqrt{6})}\approx0.204$, and the outer solution $n_{out}=const=n^\infty$ holds. On the right, $\lambda'\delta^2/L=\sqrt{{3}/{8}}\approx0.612$, and the outer solution $n_{out}=const=n^\infty$ is invalid. The left case is in the $O(\delta/L)^2$ regime (Figure~\ref{fig:Fx}), while the right case is in the linear regime.}
	\label{fig:denmap}
\end{figure}

The body shape we have chosen is shown in Figure~\ref{fig:denmap}. We purposely constructed the body shape out of four circular arcs to simplify the algorithm and to minimize the numerical error in the contact detection process in the simulations. For each combination of $\delta$ and $L$, we vary $\ell/\delta \in (1,10)$ to cover both cases $\delta\approx\ell$ and $\delta\ll\ell$.

\begin{figure}[h]
	\centering
	\includegraphics[width=\linewidth]{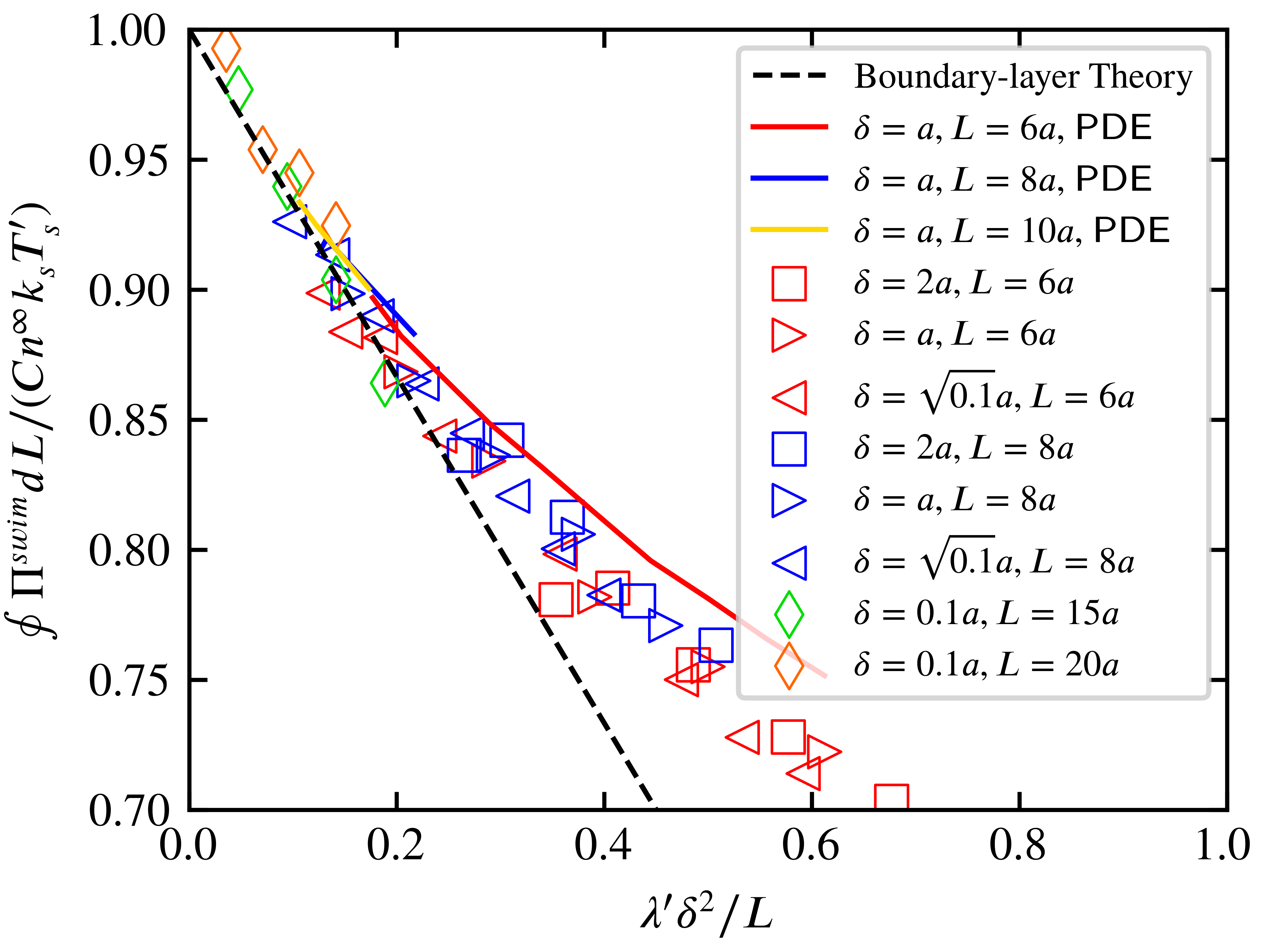}
	\caption{The scalar integration of pressure $\oint\Pi^{swim}dL$ on an asymmetric body immersed in ABPs. The dashed line is equation~(\ref{eq:ptot2Dbanana}). The symbols are simulation results for $N=40000$ particles equilibrated for $3000\tau_R$, and the solid lines are Finite Element solution for up to $10^4\tau_R$ to ensure a steady state is reached. The simulations cover the range $\ell/\delta\in[0.5,40]$, and the PDE solutions cover the range $\ell/\delta\in[0.2,5]$.}
	\label{fig:Ptot}
\end{figure}

Here we choose the inner radius of the shape as the macroscopic length scale $L$, and for the shape we used the circumference $C=3\pi L$. Equation~(\ref{eq:ptot2Dcircum}) for the integral of the pressure becomes a straight line for this shape:
\begin{align}\label{eq:ptot2Dbanana}
\frac{\oint \Pi^{swim} dL }{ n^\infty k_s T_s'C} =1-\frac{2}{3}\lambda'\delta^2/L\,. 
\end{align}
Figure~\ref{fig:Ptot} compares~(\ref{eq:ptot2Dbanana}) with simulation results and the Finite Element PDE solutions.
The theoretical expression applies for $\delta \ll L$ and $\ell \ll L$ and works well in the limit $\lambda'\delta^2/L \ll 1$ as seen in Figure~\ref{fig:Ptot}.

\begin{figure}[h]
	\centering
	\includegraphics[width=\linewidth]{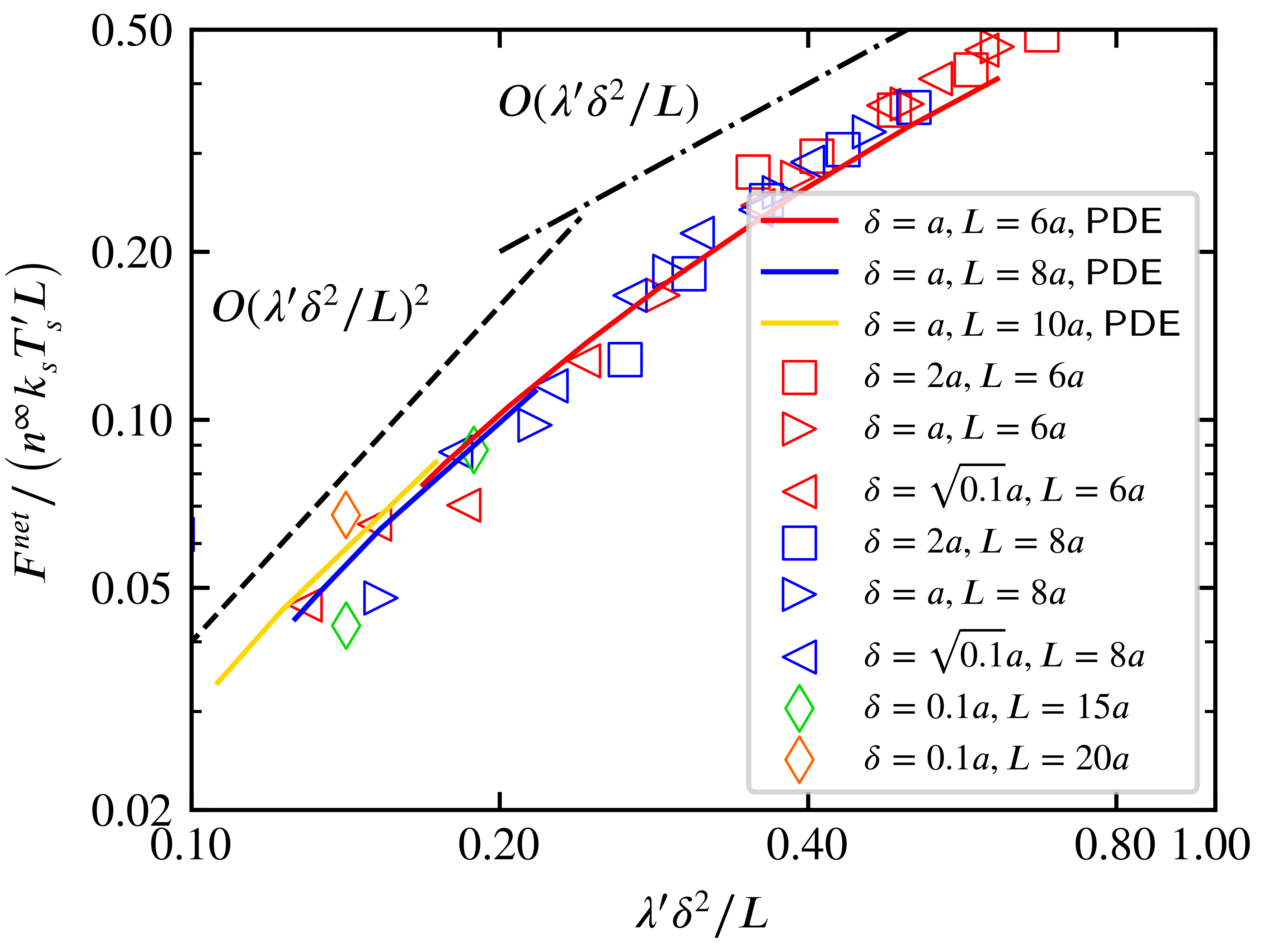}
	\caption{The net force $\bF^{net}$ on an asymmetric body immersed in ABPs. The body shape is shown in~(\ref{fig:denmap}). The data is collected from the same simulations and PDE solutions as in Figures~\ref{fig:denmap} and \ref{fig:Ptot}. The $O(\lambda^\prime \delta^2/L)^2$ and $O(\lambda^\prime \delta^2/L)$ asymptotic lines are for ease of view.}
	\label{fig:Fx}
\end{figure}

Next, we compare the theoretical estimate of the net force with the simulation results in Figure~\ref{fig:Fx}. From (\ref{eq:piswimjs2D}) and the definition of the sign of the curvature in Figure~\ref{fig:geocoordinate}, a dent in the body would increase the swim pressure while a bump on the body would decrease it. Therefore, the shape shown in Figure~\ref{fig:Fx} experiences a net force to the right. Because $\bF^{net}$ vanishes to the leading order in $\lambda'\delta^2/L$, asymptoticly a quadratic scaling emerges: $\bF^{net}\sim O(\lambda'\delta^2/L)^2$. This is verified by the results shown in Figure~\ref{fig:Fx}. Note that the simulations and numerical solution show that the data collapse according to the predictions of the boundary-layer theory even when ${\lambda'\delta^2}/L$ is not small.  That is, 
\begin{align} \label{eq:fnetscaling}
	\frac{F^{net}}{n^\infty k_sT_s L} = f\left(\frac{\lambda'\delta^2}{L}\right)\,,
\end{align}
where $f(x)$ is a function determined by the body shape, satisfying $f(x\to 0)\to x^2$. When ${\lambda'\delta^2}/{L}$ is large, $f(x)$ transitions to a linear function.

In Figure~\ref{fig:denmap}, the different cases for $\lambda'\delta^2/L\to0$ and $\lambda'\delta^2/L\to1$ are shown. It is clear that in the limit $\lambda'\delta^2/L\to 0$, our boundary layer solution is valid. In the other limit the number density $n$ still has a boundary layer on the microscopic length $\delta$ close to the boundary, but the outer solution $n_{out}=const$ is no longer valid. There is a clear wake (low density region) close to the concave portion of the body.

The wake may look contradictory with our previous argument that a dent \emph{increases} the number density, but in fact the wake is in the outer region, while the argument about the dent is for the inner region of the boundary layer. The formation of a wake with $n_{top}<n^{\infty}$ can be explained by a simple geometric argument. When $\ell \sim L$, swimmers will `pass-by' the body without reorienting into the dent;  only those swimmers with a proper orientation $\bq$ can enter the dent region. Thus the solution $n_{out}=const=n^\infty$ for $\ell\ll L$ in the analytical boundary-layer solution is no longer valid.

More precisely, our argument about the increase of $n$ at a dent is that in the inner region  the ratio $n_{wall}/n_{top}$ at a dent is higher than at a flat wall, if $n_{top}$ and $\bmm_{top}$ are the same for the dent and the flat wall. When $\ell\ll L$, we have $n_{top}=n^\infty$ and~(\ref{eq:nwallcurvature}) holds. While when $\ell \sim L$, the outer solution is no longer a constant and $n_{top}\neq n^\infty$, but our data shows that a dent still increases $n_{wall}/n_{top}$ over the flat wall value $1+\ell^2/(6\delta^2)$, although quantitatively the leading order solution~(\ref{eq:nwallcurvature}) is no longer applicable. In sum, for the wake region in the right case in Figure~\ref{fig:denmap}, we have $n_{wall}/n_{top} > 1+\ell^2/(6\delta^2)$ and $n_{top} < n^\infty$. The effect of local curvature is stronger and we still have $n_{wall}/n^\infty > 1+\ell^2/(6\delta^2)$, and therefore we have a net force  to the right.

In the limit $\lambda'\delta^2/L\to 1$, one can still take the integral of $\Pi_{wall}=n_{wall} \zeta D_T$ over the surface to get the net force, but it is no longer correct to use the constant outer solution $n_{out}=const=n^\infty$. In this case the global transport equations~(\ref{eq:jn}) must be solved to get the correct number density field from which to calculate the force, as we did in Figure~\ref{fig:Fx} with the PDE solver. Physically, when $\ell\sim L$, in a single run-length a swimmer does not sample only the local geometry, but actually experiences the global body shape (cf. Figure~\ref{fig:inout}). Therefore, our solution of a completely localized boundary layer is no longer valid.

\section{Discussion \& Conclusions}
In this paper we analyzed the kinetic accumulation boundary layer of ABPs for the general case where $\delta/L$ and $\ell/L$ are both small, but the ratio $\delta/\ell$ is arbitrary. We found that a universal scaling emerges with $\lambda\delta^2/L$ as the governing parameter, and our analytical solution for the boundary layer is valid for $\lambda\delta^2/L\ll 1$. When $\lambda\delta^2/L$ is not small, we showed by simulations and PDE solutions that the scaling and the boundary layer structure still holds, but the outer solution $n_{out}$ is no longer a constant $n^\infty$. Also, in the analytical and PDE solutions the moment expansion is truncated at the polar order, $\bmm$, level with an assumption of negligible nematic order, $\bQ = 0$. The solution matches the Brownian simulations well, as was the case discussed in detail in \citet{ForceBoundary2015}.

While we discussed only the exterior problem, the boundary-layer structure and solution is not limited to this case. It is also applicable to an interior problem where swimmers are confined in an arbitrary shaped container and form a boundary layer on the interior walls.

In many experiments, $\delta \ll L$ and interest is for fast swimmers where $\ell\gg\delta$. In the following, we shall discuss the behavior for $\ell$ relative to the macroscopic length $L$.

The disappearance of a net force at first order in $\lambda\delta^2/L$ is due to the fact that the outer solution is constant: $n_{out}=n^\infty=const$. Since the outer solution is governed by Laplace's equation, the constant solution is determined by the no-flux boundary condition: $j_{n,\perp}=0$. The vanishing of the normal component of the flux in the outer regions arises from the continuity of the flux inside the boundary layer and the no flux condition at the actual body surface. 

In the boundary layer we expanded the solution at the surface with a geometric constant $J_S$. But $J_S$ is only the mean curvature and it does not take into consideration any variation of the curvature. The surface has a \emph{constant} curvature, which  means that  the local solution is the same at all points along the surface. Thus, there is no tangential flux of active particles, and by the continuity of flux across the boundary layer, there is no flux into or out of the boundary-layer region.

\begin{figure}
	\centering
	\includegraphics[width=\linewidth]{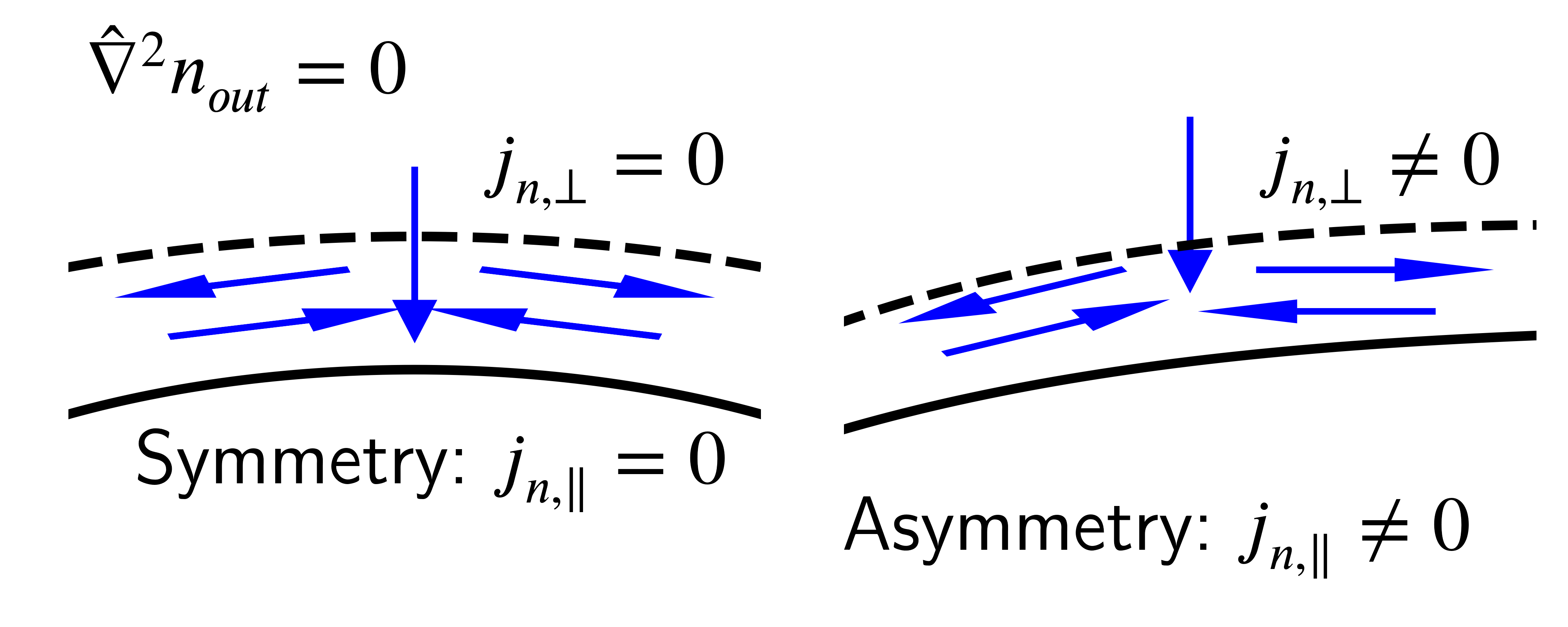}
	\caption{The origin of a non zero flux $j_{n,\perp}$. In the left, curvature is symmetric. On the right, the curvature is asymmetric.}
	\label{fig:inout}
\end{figure}

If we allow the curvature to vary along the surface, however, then a non-zero tangential flux is possible as illustrated in Figure~\ref{fig:inout}. By continuity, a normal flux from the outer region can come into the boundary layer and flow out tangentially along the surface. A non-zero normal flux $j_{n,\perp}$ will give rise to a non-constant outer solution $n_{out}$. When the boundary-layer assumption holds, the variation in curvature appears at second order in the expansion of the curvilinear operator $\nablas$, as discussed in the Appendix. The departure of $n$ from a constant outer solution $n^\infty$ thus occurs at second order, at $O(\lambda\delta^2/L)^2$, and therefore we observe a second order net force as shown in~(\ref{eq:fnetscaling}) and in Figure~\ref{fig:Fx}. An analytical theory for the second order would require all 21 geometric constants for curvature and curvature variations to be included and is overly complicated.\cite{Panaras1987}

\subsection{$\ell\ll L$: $\bF^{net}\sim O(\lambda\delta^2/L)^2 n^\infty k_sT_s L^2$.}

When $\delta\ll\ell$, $\lambda\delta^2/L \sim \ell/L$, which corresponds to the singular limit of no translational Brownian diffusion, $D_T = 0$, and explains the appearance of the $\ell/L$ scalings in literature.\cite{FilyConfinedSwimmers2014,Smallenburg2015} However, it is important to note that in this case the boundary-layer thickness goes to zero and one cannot simply put the Smoluchowski equation into a PDE solver with a finite minimum mesh size $l_m$. In fact, in this case the number density $n$ behaves as a Dirac delta function at the boundary and cannot be properly captured by any finite mesh size. The proper procedure for zero $\delta$ is to  split the particles into a surface layer `on the wall' and a bulk distribution to capture the number density field.\cite{Ezhilan2015}  If one blindly chooses some $l_m$, it is equivalent to specifying a finite microscopic length $\delta$, which may lead to a spurious constant in front of the  $\ell/L$ scaling. Also, only the linear leading order correction dependent on $J_S$ can be attached to a constant outer solution. It is not legitimate to attach a numerically fitted high order inner solution with a leading order constant outer solution to form the surface integral of the net force.

\subsection{$\ell\sim L$: $\bF^{net}\sim O(\lambda\delta^2/L)n^\infty k_sT_s L^2 \sim O(\ell/L) n^\infty k_sT_s L^2 $}
In this case since we assumed $\delta\ll L$, we have $F^{net}\sim O(\ell/L)nk_sT_sL^2$. As shown in Figure~\ref{fig:denmap} there is also a boundary layer governed by the microscopic length $\delta$, but on the large scale of $L$ the outer solution shows a clear advection-like wake structure. In contrast to the case where $\ell \ll L$, the swimmers explore the variation in the curvature over the entire body within a single run and cause the departure of $n_{out}$ from a constant $n^\infty$. More specifically, in this case only a few swimmers with preferred orientation $\bq$ can enter the concave portion of the macroscopic body, and therefore a long-range orientation field $\bmm_{out} = -\tfrac{\ell}{6L} \hnabla n_{out}$ appears together with a non constant $n_{out}$. The surface integral, (\ref{eq:integral3D}), for $\bF^{net}$ is now $\oint n_{out} J_S\bn d\hat{L}$ and the variation of $n_{out}$ along the surface gives rise to net force linearly dependent on $\lambda\delta^2/L$.

Also in this limit for a regular shaped circle or sphere the full exact  solution\cite{ForceBoundary2015} shows that the Pad\'{e} form:
\begin{align}
	\Pi_{wall} \to n^\infty k_BT + n^\infty k_sT_s \frac{1}{1 - \frac{1}{2\sqrt{3}}\frac{\ell}{L} \left(\hkappa_1 + \hkappa_2\right)}\,,
\end{align}
is the exact analytic solution and works well even when $\ell/L\sim 5$. The linear expansion of this Pad\'{e} form is the first order boundary layer solution~(\ref{eq:pi3dellL}). However, we cannot write~(\ref{eq:piswimjs}) into this Pad\'{e} form and attach it to the constant outer solution $n_{out}=n^{\infty}$ to calculate the force, because the force inherently requires a second order outer solution.

The difference between the limits $\ell\ll L$ and $\ell \sim L$ can also be appreciated from a continuum mechanics point of view. We have shown that for swimmers with no orientational bias or body force on large scales, $L\gg\ell$,  continuum mechanics describes the number density flux well\cite{SwimForce2015} and
\begin{align}\label{eq:jnstress}
	\bj_n = -\frac{1}{\zeta}\nabla\cdot\bsigma^{act}\,,
\end{align}
where, if the swimmer-swimmer interaction is ignored (the dilute limit), the active stress is
\begin{align}
	\bsigma^{act} = - n \zeta \left(D_T + D^{swim} \right)\bI\,.
\end{align}
In this continuum formulation, any non-continuum effects are only important in a very thin layer attached to the body surface, which is the boundary layer considered in this paper. The continuum mechanical flux~(\ref{eq:jnstress}) is exactly the outer flux~(\ref{eq:jnouter}). Our non-dimensional scale factor  $\lambda\delta^2/L$ is the counterpart to the Knudsen number in rarefied gas dynamics:  the ratio of the mean free path to body size $Kn=\lambda_{MFP}/L$. When $Kn\lesssim 0.1$ the Navier-Stokes equation is applicable in the bulk, but with a boundary condition modified by the Knudsen layer close to a surface. In our solution, Figures~\ref{fig:Ptot} and \ref{fig:Fx}, we also see that the first order boundary-layer solution is valid when $\lambda\delta^2/L \lesssim 0.2$.  However, when $\ell\sim L$, the continuum mechanics transport equation can only be used in regions  far away from the body, and we need to solve the detailed Smoluchowski equation in the vicinity of the body. This is similar to the transition regime $Kn\approx 1$ in rarefied gas dynamics  where the detailed dynamics must be considered.

\subsection{General solution: spherical harmonics}
The decoupled structure~(\ref{eq:mdivscreen}) allows us to construct the general solution for $f=\hnabla\cdot\bmm$  by spherical harmonics (similar in 2D), and then match the boundary conditions to get the general solution without going into the details of the curvature expansion. However, in the case where $\lambda\delta^2/L\ll 1$, numerically the series constants are highly sensitive to numerical errors, and for a body of complex shape the general solution is of little utility.  Here, we present the general structure for an axisymmetric body to complete our mathematical discussion.

For a general 3D  body  axisymmetric about the $z$-axis:
\begin{align}
\hnabla\cdot\bmm &= \sum_{l=0}^{\infty} B_{l} \sqrt{\frac{2}{\pi \lambda L r}} K_{l+\frac{1}{2}}(\lambda L r) P_l(\cos\theta)\,, \\
\begin{split}
n&=n_H+n_P\\
&= n^\infty+ \sum_{l=0}^{\infty} g_l r^{-(l+1)} P_l(\cos\theta) + \frac{\ell}{\delta^2\lambda^2 L} \hnabla\cdot\bmm\,,
\end{split}\\
\begin{split}
\bmm &= \bmm_H + \bmm_P\\
&= \sum_{l=0}^{\infty} \bC_{l}(\theta) \sqrt{\frac{2}{\pi \alpha r}} K_{l+\frac{1}{2}}(\alpha r) \\
& + \frac{1}{\lambda^2L^2} \hnabla(\hnabla\cdot\bmm) - \frac{l}{6L} \hnabla n_H\,,
\end{split}
\end{align}
where $K_{l+\frac{1}{2}}(z)$ are {\em{cylindrical}} modified Bessel functions. For interior problems, $K_{l+\frac{1}{2}}(z)$ should be replaced by $I_l(z)$. $P_l(x)$ is the Legendre polynomial and $B_l,g_l,\bC_l$ are determined  by the no-flux boundary condition.

For a general body in 2D,  we can replace the Legendre polynomials by Fourier modes $\cos l\theta + \sin l\theta$, and replace the $K_{l+\frac{1}{2}}(z)$ with its integer order version $K_l(z)$, to construct the general solution. The constants should also be adjusted accordingly.

\subsection{Including hydrodynamics}
In this paper we discussed the kinetic limit where the boundary layer emerges solely due to the run length $\ell=U_0\tau_R$ and the microscopic length $\delta=\sqrt{D_T\tau_R}$. We showed that the interaction is completely determined by the distribution of swimmers around the body---the distribution function $P(\bx,\bq,t)$. 

This is also true for swimmers with full hydrodynamics in Stokes flow. As shown by the equation~(4.19) in \citet{Brady2011}, the hydrodynamic force applied on the body is completely determined by the distribution function  $p_{ij} (\br)$ of a swimmer $j$ relative to a macroscopic body $i$, and the full mobility matrix $\bM_{ij}(\br)$.

In the limit where the excluded volume interaction is effective on a range much longer than hydrodynamic interactions, the hydrodynamic mobility is simplified to the isotropic Stokes drag, and the interaction is simplified to the sum of individual Brownian collisions as discussed in this paper.  When hydrodynamic interactions are important, the distribution function $p_{ij}(\br)$ must be solved to find the correct interaction and boundary-layer form. We leave this for a future study.

\section{Acknowledgment}
We thank Eric W. Burkholder for the mathematical construction~(\ref{eq:mdivscreen}). This work is supported by NSF-CBET 1437570.

\section{Appendix: principal curvatures}\label{sec:appendix}
For a curve on a 2D plane, described by a parametrized curve $(x(t),y(t))$, the curvature is well-known:
\begin{align}
	\kappa = - \frac{\ABS{x'y''-x''y'}}{\left(x'^2+y'^2\right)^{3/2}}\,.
\end{align}

For a smooth surface in 3D space at any point the planes of two principal curvatures and the tangent plane are perpendicular to each other.\cite{Brenner2013}
The mean curvature non-dimensionalized by $L$ is simply determined by the surface normal vector $\bn$:
\begin{align}
	H=\frac{\hkappa_1+\hkappa_2}{2} = -\frac{1}{2}\hnabla\cdot\bn\,.
\end{align} 

The details of how to build a curvilinear coordinate system and all the expansion of operators can be found in the Appendix of the work by \citet{Brenner2013}. Without going to the tedious algebraic details, here we only include the relevant leading order expansion of the operators:
	\begin{subequations}\label{eq:nablaGeo}
		\begin{align}
			\nablas f =& \dpone{f}{q_\perp} \bn + O\left(\frac{\delta}{L}\right)^2\,,\\
			\nablas\cdot \bg = &  - J_S g_\perp + \dpone{g_\perp}{q_\perp}+O\left(\frac{\delta}{L}\right)^2\,,\\
			\nablas^2 f =& - J_S \dpone{f}{q_\perp} + \dptwo{f}{q_\perp}+O\left(\frac{\delta}{L}\right)^2\,,
		\end{align}
	\end{subequations}
where $J_S=2H$, $\bg=(0,0,g_\perp)$, and $\nablas$ is the nabla operator in the (orthogonal) curvilinear coordinate system $q_1,q_2,q_\perp$, defined on the curved surface shown in Figure~\ref{fig:geocoordinate}. $q_\perp$ follows the direction of the surface normal vector $\bn$. $q_1$ and $q_2$ are on the curved surface and are located in the two planes associated with the two principal vectors, respectively. It is clear that to leading order $O(\delta/L)$, there is no need to deal with the gradients in the  $q_1,q_2$ directions. The formulation of~(\ref{eq:nablaGeo}) is consistent with the work in literature\cite{Cox1997,Yariv2009} on curved boundary layers. 

The expansion of operator $\nablas$ beyond the leading order relies on a rigorous algebra of the full curvilinear space. Mathematically, the second order expansion relies not only on the mean curvature $H$, but also on the {\em variations} of curvatures. The full expansion may include up to 21 curvature coefficients,\cite{Panaras1987} and probably forbids any analytical work.





\bibliography{ref} 

\begin{thebibliography}{23}%
\makeatletter
\providecommand \@ifxundefined [1]{%
 \@ifx{#1\undefined}
}%
\providecommand \@ifnum [1]{%
 \ifnum #1\expandafter \@firstoftwo
 \else \expandafter \@secondoftwo
 \fi
}%
\providecommand \@ifx [1]{%
 \ifx #1\expandafter \@firstoftwo
 \else \expandafter \@secondoftwo
 \fi
}%
\providecommand \natexlab [1]{#1}%
\providecommand \enquote  [1]{``#1''}%
\providecommand \bibnamefont  [1]{#1}%
\providecommand \bibfnamefont [1]{#1}%
\providecommand \citenamefont [1]{#1}%
\providecommand \href@noop [0]{\@secondoftwo}%
\providecommand \href [0]{\begingroup \@sanitize@url \@href}%
\providecommand \@href[1]{\@@startlink{#1}\@@href}%
\providecommand \@@href[1]{\endgroup#1\@@endlink}%
\providecommand \@sanitize@url [0]{\catcode `\\12\catcode `\$12\catcode
  `\&12\catcode `\#12\catcode `\^12\catcode `\_12\catcode `\%12\relax}%
\providecommand \@@startlink[1]{}%
\providecommand \@@endlink[0]{}%
\providecommand \url  [0]{\begingroup\@sanitize@url \@url }%
\providecommand \@url [1]{\endgroup\@href {#1}{\urlprefix }}%
\providecommand \urlprefix  [0]{URL }%
\providecommand \Eprint [0]{\href }%
\providecommand \doibase [0]{http://dx.doi.org/}%
\providecommand \selectlanguage [0]{\@gobble}%
\providecommand \bibinfo  [0]{\@secondoftwo}%
\providecommand \bibfield  [0]{\@secondoftwo}%
\providecommand \translation [1]{[#1]}%
\providecommand \BibitemOpen [0]{}%
\providecommand \bibitemStop [0]{}%
\providecommand \bibitemNoStop [0]{.\EOS\space}%
\providecommand \EOS [0]{\spacefactor3000\relax}%
\providecommand \BibitemShut  [1]{\csname bibitem#1\endcsname}%
\let\auto@bib@innerbib\@empty
\bibitem [{\citenamefont {Berke}\ \emph {et~al.}(2008)\citenamefont {Berke},
  \citenamefont {Turner}, \citenamefont {Berg},\ and\ \citenamefont
  {Lauga}}]{Berke2008}%
  \BibitemOpen
  \bibfield  {author} {\bibinfo {author} {\bibfnamefont {A.~P.}\ \bibnamefont
  {Berke}}, \bibinfo {author} {\bibfnamefont {L.}~\bibnamefont {Turner}},
  \bibinfo {author} {\bibfnamefont {H.~C.}\ \bibnamefont {Berg}}, \ and\
  \bibinfo {author} {\bibfnamefont {E.}~\bibnamefont {Lauga}},\ }\href
  {\doibase 10.1103/PhysRevLett.101.038102} {\bibfield  {journal} {\bibinfo
  {journal} {Phys. Rev. Lett.}\ }\textbf {\bibinfo {volume} {101}},\ \bibinfo
  {pages} {{038102}} (\bibinfo {year} {2008})}\BibitemShut {NoStop}%
\bibitem [{\citenamefont {Li}\ and\ \citenamefont {Tang}(2009)}]{Li2009}%
  \BibitemOpen
  \bibfield  {author} {\bibinfo {author} {\bibfnamefont {G.}~\bibnamefont
  {Li}}\ and\ \bibinfo {author} {\bibfnamefont {J.~X.}\ \bibnamefont {Tang}},\
  }\href {\doibase 10.1103/PhysRevLett.103.078101} {\bibfield  {journal}
  {\bibinfo  {journal} {Phys. Rev. Lett.}\ }\textbf {\bibinfo {volume} {103}},\
  \bibinfo {pages} {{078101}} (\bibinfo {year} {2009})}\BibitemShut {NoStop}%
\bibitem [{\citenamefont {Li}\ \emph {et~al.}(2011)\citenamefont {Li},
  \citenamefont {Bensson}, \citenamefont {Nisimova}, \citenamefont {Munger},
  \citenamefont {Mahautmr}, \citenamefont {Tang}, \citenamefont {Maxey},\ and\
  \citenamefont {Brun}}]{Li2011}%
  \BibitemOpen
  \bibfield  {author} {\bibinfo {author} {\bibfnamefont {G.}~\bibnamefont
  {Li}}, \bibinfo {author} {\bibfnamefont {J.}~\bibnamefont {Bensson}},
  \bibinfo {author} {\bibfnamefont {L.}~\bibnamefont {Nisimova}}, \bibinfo
  {author} {\bibfnamefont {D.}~\bibnamefont {Munger}}, \bibinfo {author}
  {\bibfnamefont {P.}~\bibnamefont {Mahautmr}}, \bibinfo {author}
  {\bibfnamefont {J.~X.}\ \bibnamefont {Tang}}, \bibinfo {author}
  {\bibfnamefont {M.~R.}\ \bibnamefont {Maxey}}, \ and\ \bibinfo {author}
  {\bibfnamefont {Y.~V.}\ \bibnamefont {Brun}},\ }\href {\doibase
  10.1103/PhysRevE.84.041932} {\bibfield  {journal} {\bibinfo  {journal} {Phys.
  Rev. E}\ }\textbf {\bibinfo {volume} {84}},\ \bibinfo {pages} {{041932}}
  (\bibinfo {year} {2011})}\BibitemShut {NoStop}%
\bibitem [{\citenamefont {Uspal}\ \emph {et~al.}(2015)\citenamefont {Uspal},
  \citenamefont {Popescu}, \citenamefont {Dietrich},\ and\ \citenamefont
  {Tasinkevych}}]{Uspal2015}%
  \BibitemOpen
  \bibfield  {author} {\bibinfo {author} {\bibfnamefont {W.~E.}\ \bibnamefont
  {Uspal}}, \bibinfo {author} {\bibfnamefont {M.~N.}\ \bibnamefont {Popescu}},
  \bibinfo {author} {\bibfnamefont {S.}~\bibnamefont {Dietrich}}, \ and\
  \bibinfo {author} {\bibfnamefont {M.}~\bibnamefont {Tasinkevych}},\ }\href
  {\doibase 10.1039/C5SM01088H} {\bibfield  {journal} {\bibinfo  {journal}
  {Soft Matter}\ }\textbf {\bibinfo {volume} {11}},\ \bibinfo {pages} {6613}
  (\bibinfo {year} {2015})}\BibitemShut {NoStop}%
\bibitem [{\citenamefont {Kaya}\ and\ \citenamefont {Koser}(2012)}]{Kaya2012}%
  \BibitemOpen
  \bibfield  {author} {\bibinfo {author} {\bibfnamefont {T.}~\bibnamefont
  {Kaya}}\ and\ \bibinfo {author} {\bibfnamefont {H.}~\bibnamefont {Koser}},\
  }\href {\doibase 10.1016/j.bpj.2012.03.001} {\bibfield  {journal} {\bibinfo
  {journal} {Biophys. J.}\ }\textbf {\bibinfo {volume} {102}},\ \bibinfo
  {pages} {1514} (\bibinfo {year} {2012})}\BibitemShut {NoStop}%
\bibitem [{\citenamefont {Lauga}\ \emph {et~al.}(2006)\citenamefont {Lauga},
  \citenamefont {DiLuzio}, \citenamefont {Whitesides},\ and\ \citenamefont
  {Stone}}]{Lauga2006}%
  \BibitemOpen
  \bibfield  {author} {\bibinfo {author} {\bibfnamefont {E.}~\bibnamefont
  {Lauga}}, \bibinfo {author} {\bibfnamefont {W.~R.}\ \bibnamefont {DiLuzio}},
  \bibinfo {author} {\bibfnamefont {G.~M.}\ \bibnamefont {Whitesides}}, \ and\
  \bibinfo {author} {\bibfnamefont {H.~A.}\ \bibnamefont {Stone}},\ }\href
  {\doibase 10.1529/biophysj.105.069401} {\bibfield  {journal} {\bibinfo
  {journal} {Biophys. J.}\ }\textbf {\bibinfo {volume} {90}},\ \bibinfo {pages}
  {400} (\bibinfo {year} {2006})}\BibitemShut {NoStop}%
\bibitem [{\citenamefont {Kaiser}\ \emph {et~al.}(2014)\citenamefont {Kaiser},
  \citenamefont {Peshkov}, \citenamefont {Sokolov}, \citenamefont {ten Hagen},
  \citenamefont {L{\"o}wen},\ and\ \citenamefont {Aranson}}]{Kaiser2014}%
  \BibitemOpen
  \bibfield  {author} {\bibinfo {author} {\bibfnamefont {A.}~\bibnamefont
  {Kaiser}}, \bibinfo {author} {\bibfnamefont {A.}~\bibnamefont {Peshkov}},
  \bibinfo {author} {\bibfnamefont {A.}~\bibnamefont {Sokolov}}, \bibinfo
  {author} {\bibfnamefont {B.}~\bibnamefont {ten Hagen}}, \bibinfo {author}
  {\bibfnamefont {H.}~\bibnamefont {L{\"o}wen}}, \ and\ \bibinfo {author}
  {\bibfnamefont {I.~S.}\ \bibnamefont {Aranson}},\ }\href {\doibase
  10.1103/PhysRevLett.112.158101} {\bibfield  {journal} {\bibinfo  {journal}
  {Phys. Rev. Lett.}\ }\textbf {\bibinfo {volume} {112}},\ \bibinfo {pages}
  {158101} (\bibinfo {year} {2014})}\BibitemShut {NoStop}%
\bibitem [{\citenamefont {Yan}\ and\ \citenamefont
  {Brady}(2015{\natexlab{a}})}]{SwimForce2015}%
  \BibitemOpen
  \bibfield  {author} {\bibinfo {author} {\bibfnamefont {W.}~\bibnamefont
  {Yan}}\ and\ \bibinfo {author} {\bibfnamefont {J.~F.}\ \bibnamefont
  {Brady}},\ }\href {\doibase 10.1039/C5SM01318F} {\bibfield  {journal}
  {\bibinfo  {journal} {Soft Matter}\ }\textbf {\bibinfo {volume} {11}},\
  \bibinfo {pages} {6235} (\bibinfo {year} {2015}{\natexlab{a}})}\BibitemShut
  {NoStop}%
\bibitem [{\citenamefont {Ezhilan}\ \emph {et~al.}(2015)\citenamefont
  {Ezhilan}, \citenamefont {Alonso-Matilla},\ and\ \citenamefont
  {Saintillan}}]{Ezhilan2015}%
  \BibitemOpen
  \bibfield  {author} {\bibinfo {author} {\bibfnamefont {B.}~\bibnamefont
  {Ezhilan}}, \bibinfo {author} {\bibfnamefont {R.}~\bibnamefont
  {Alonso-Matilla}}, \ and\ \bibinfo {author} {\bibfnamefont {D.}~\bibnamefont
  {Saintillan}},\ }\href {\doibase 10.1017/jfm.2015.520} {\bibfield  {journal}
  {\bibinfo  {journal} {J. Fluid Mech.}\ }\textbf {\bibinfo {volume} {781}},\
  \bibinfo {pages} {R4} (\bibinfo {year} {2015})}\BibitemShut {NoStop}%
\bibitem [{\citenamefont {Takatori}\ \emph {et~al.}(2014)\citenamefont
  {Takatori}, \citenamefont {Yan},\ and\ \citenamefont {Brady}}]{Pressure2014}%
  \BibitemOpen
  \bibfield  {author} {\bibinfo {author} {\bibfnamefont {S.~C.}\ \bibnamefont
  {Takatori}}, \bibinfo {author} {\bibfnamefont {W.}~\bibnamefont {Yan}}, \
  and\ \bibinfo {author} {\bibfnamefont {J.~F.}\ \bibnamefont {Brady}},\ }\href
  {\doibase 10.1103/PhysRevLett.113.028103} {\bibfield  {journal} {\bibinfo
  {journal} {Phys. Rev. Lett.}\ }\textbf {\bibinfo {volume} {113}},\ \bibinfo
  {pages} {{028103}} (\bibinfo {year} {2014})}\BibitemShut {NoStop}%
\bibitem [{\citenamefont {Yan}\ and\ \citenamefont
  {Brady}(2015{\natexlab{b}})}]{ForceBoundary2015}%
  \BibitemOpen
  \bibfield  {author} {\bibinfo {author} {\bibfnamefont {W.}~\bibnamefont
  {Yan}}\ and\ \bibinfo {author} {\bibfnamefont {J.~F.}\ \bibnamefont
  {Brady}},\ }\href {\doibase 10.1017/jfm.2015.621} {\bibfield  {journal}
  {\bibinfo  {journal} {J. Fluid Mech.}\ }\textbf {\bibinfo {volume} {785}},\
  \bibinfo {pages} {R1} (\bibinfo {year} {2015}{\natexlab{b}})}\BibitemShut
  {NoStop}%
\bibitem [{\citenamefont {Takatori}\ and\ \citenamefont
  {Brady}(2016)}]{takatori_forces_2016}%
  \BibitemOpen
  \bibfield  {author} {\bibinfo {author} {\bibfnamefont {S.~C.}\ \bibnamefont
  {Takatori}}\ and\ \bibinfo {author} {\bibfnamefont {J.~F.}\ \bibnamefont
  {Brady}},\ }\href {\doibase 10.1016/j.cocis.2015.12.003} {\bibfield
  {journal} {\bibinfo  {journal} {Current Opinion in Colloid \& Interface
  Science}\ }\textbf {\bibinfo {volume} {21}},\ \bibinfo {pages} {24} (\bibinfo
  {year} {2016})}\BibitemShut {NoStop}%
\bibitem [{\citenamefont {Saintillan}\ and\ \citenamefont
  {Shelley}(2015)}]{Saintillan2015}%
  \BibitemOpen
  \bibfield  {author} {\bibinfo {author} {\bibfnamefont {D.}~\bibnamefont
  {Saintillan}}\ and\ \bibinfo {author} {\bibfnamefont {M.~J.}\ \bibnamefont
  {Shelley}},\ }in\ \href {\doibase 10.1007/978-1-4939-2065-5_9} {\emph
  {\bibinfo {booktitle} {Complex Fluids in Biological Systems}}},\ \bibinfo
  {series and number} {Biological and Medical Physics, Biomedical
  Engineering},\ \bibinfo {editor} {edited by\ \bibinfo {editor} {\bibfnamefont
  {S.~E.}\ \bibnamefont {Spagnolie}}}\ (\bibinfo  {publisher} {Springer New
  York},\ \bibinfo {year} {2015})\ Chap.~\bibinfo {chapter} {9}, pp.\ \bibinfo
  {pages} {319--355}\BibitemShut {NoStop}%
\bibitem [{\citenamefont {Fily}\ \emph {et~al.}(2014)\citenamefont {Fily},
  \citenamefont {Baskaran},\ and\ \citenamefont
  {Hagan}}]{FilyConfinedSwimmers2014}%
  \BibitemOpen
  \bibfield  {author} {\bibinfo {author} {\bibfnamefont {Y.}~\bibnamefont
  {Fily}}, \bibinfo {author} {\bibfnamefont {A.}~\bibnamefont {Baskaran}}, \
  and\ \bibinfo {author} {\bibfnamefont {M.~F.}\ \bibnamefont {Hagan}},\ }\href
  {\doibase 10.1039/C4SM00975D} {\bibfield  {journal} {\bibinfo  {journal}
  {Soft Matter}\ }\textbf {\bibinfo {volume} {10}},\ \bibinfo {pages} {5609}
  (\bibinfo {year} {2014})}\BibitemShut {NoStop}%
\bibitem [{\citenamefont {Smallenburg}\ and\ \citenamefont
  {L\"{o}wen}(2015)}]{Smallenburg2015}%
  \BibitemOpen
  \bibfield  {author} {\bibinfo {author} {\bibfnamefont {F.}~\bibnamefont
  {Smallenburg}}\ and\ \bibinfo {author} {\bibfnamefont {H.}~\bibnamefont
  {L\"{o}wen}},\ }\href {\doibase 10.1103/PhysRevE.92.032304} {\bibfield
  {journal} {\bibinfo  {journal} {Phys. Rev. E}\ }\textbf {\bibinfo {volume}
  {92}},\ \bibinfo {pages} {{032304}} (\bibinfo {year} {2015})}\BibitemShut
  {NoStop}%
\bibitem [{\citenamefont {Brady}(2017)}]{Brady2017}%
  \BibitemOpen
  \bibfield  {author} {\bibinfo {author} {\bibfnamefont {J.}~\bibnamefont
  {Brady}},\ }\href@noop {} {\bibfield  {journal} {\bibinfo  {journal} {In
  preparation}\ } (\bibinfo {year} {2017})}\BibitemShut {NoStop}%
\bibitem [{\citenamefont {Brenner}\ and\ \citenamefont
  {Condiff}(1972)}]{Brenner1972}%
  \BibitemOpen
  \bibfield  {author} {\bibinfo {author} {\bibfnamefont {H.}~\bibnamefont
  {Brenner}}\ and\ \bibinfo {author} {\bibfnamefont {D.~W.}\ \bibnamefont
  {Condiff}},\ }\href {\doibase 10.1016/0021-9797(72)90111-7} {\bibfield
  {journal} {\bibinfo  {journal} {J. Colloid Interface Sci.}\ }\textbf
  {\bibinfo {volume} {41}},\ \bibinfo {pages} {228} (\bibinfo {year}
  {1972})}\BibitemShut {NoStop}%
\bibitem [{\citenamefont {Foss}\ and\ \citenamefont {Brady}(2000)}]{Foss2000}%
  \BibitemOpen
  \bibfield  {author} {\bibinfo {author} {\bibfnamefont {D.~R.}\ \bibnamefont
  {Foss}}\ and\ \bibinfo {author} {\bibfnamefont {J.~F.}\ \bibnamefont
  {Brady}},\ }\href {\doibase 10.1122/1.551104} {\bibfield  {journal} {\bibinfo
   {journal} {J. Rheol.}\ }\textbf {\bibinfo {volume} {44}},\ \bibinfo {pages}
  {629} (\bibinfo {year} {2000})}\BibitemShut {NoStop}%
\bibitem [{\citenamefont {Panaras}(1987)}]{Panaras1987}%
  \BibitemOpen
  \bibfield  {author} {\bibinfo {author} {\bibfnamefont {A.~G.}\ \bibnamefont
  {Panaras}},\ }\href@noop {} {\emph {\bibinfo {title} {Boundary-layer
  equations in generalized curvilinear coordinates}}},\ \bibinfo {type} {Tech.
  Rep.}\ \bibinfo {number} {NASA-TM-100003, A-87272, NAS 1.15:100003}\
  (\bibinfo  {institution} {NASA},\ \bibinfo {year} {1987})\BibitemShut
  {NoStop}%
\bibitem [{\citenamefont {Brady}(2011)}]{Brady2011}%
  \BibitemOpen
  \bibfield  {author} {\bibinfo {author} {\bibfnamefont {J.~F.}\ \bibnamefont
  {Brady}},\ }\href {\doibase 10.1017/S0022112010004404} {\bibfield  {journal}
  {\bibinfo  {journal} {J. Fluid Mech.}\ }\textbf {\bibinfo {volume} {667}},\
  \bibinfo {pages} {216} (\bibinfo {year} {2011})}\BibitemShut {NoStop}%
\bibitem [{\citenamefont {Edwards}\ \emph {et~al.}(2013)\citenamefont
  {Edwards}, \citenamefont {Brenner},\ and\ \citenamefont
  {Wasan}}]{Brenner2013}%
  \BibitemOpen
  \bibfield  {author} {\bibinfo {author} {\bibfnamefont {D.~A.}\ \bibnamefont
  {Edwards}}, \bibinfo {author} {\bibfnamefont {H.}~\bibnamefont {Brenner}}, \
  and\ \bibinfo {author} {\bibfnamefont {D.~T.}\ \bibnamefont {Wasan}},\
  }\href@noop {} {\emph {\bibinfo {title} {Interfacial Transport Processes and
  Rheology}}},\ Butterworth-Heinemann series in chemical engineering\ (\bibinfo
   {publisher} {Butterworth-Heinemann},\ \bibinfo {year} {2013})\BibitemShut
  {NoStop}%
\bibitem [{\citenamefont {Cox}(1997)}]{Cox1997}%
  \BibitemOpen
  \bibfield  {author} {\bibinfo {author} {\bibfnamefont {R.~G.}\ \bibnamefont
  {Cox}},\ }\href {\doibase 10.1017/S0022112097004862} {\bibfield  {journal}
  {\bibinfo  {journal} {J. Fluid Mech.}\ }\textbf {\bibinfo {volume} {338}},\
  \bibinfo {pages} {1} (\bibinfo {year} {1997})}\BibitemShut {NoStop}%
\bibitem [{\citenamefont {Yariv}(2009)}]{Yariv2009}%
  \BibitemOpen
  \bibfield  {author} {\bibinfo {author} {\bibfnamefont {E.}~\bibnamefont
  {Yariv}},\ }\href {\doibase 10.1080/00986440903076590} {\bibfield  {journal}
  {\bibinfo  {journal} {Chem. Eng. Commun.}\ }\textbf {\bibinfo {volume}
  {197}},\ \bibinfo {pages} {3} (\bibinfo {year} {2009})}\BibitemShut {NoStop}%
\end{thebibliography}

\end{document}